\PassOptionsToPackage{unicode}{hyperref}
\PassOptionsToPackage{hyphens}{url}
\PassOptionsToPackage{dvipsnames,svgnames,x11names}{xcolor}

\documentclass[
  authoryear,
  preprint,
  3p,
  onecolumn]{elsarticle}
\usepackage{microtype}
\usepackage{amsmath,amssymb}
\usepackage{iftex}
\usepackage{booktabs}
\usepackage{afterpage}

\usepackage{lineno}
\usepackage{xr}
\usepackage{xr-hyper}
\ifPDFTeX
  \usepackage[T1]{fontenc}
  \usepackage[utf8]{inputenc}
  \usepackage{textcomp} 
\else 
  \usepackage{unicode-math}
  \defaultfontfeatures{Scale=MatchLowercase}
  \defaultfontfeatures[\rmfamily]{Ligatures=TeX,Scale=1}
\fi
\usepackage{lmodern}
\ifPDFTeX\else  
\fi
\IfFileExists{upquote.sty}{\usepackage{upquote}}{}
\IfFileExists{microtype.sty}{
  \usepackage[]{microtype}
  \UseMicrotypeSet[protrusion]{basicmath} 
}{}
\makeatletter
\@ifundefined{KOMAClassName}{
  \IfFileExists{parskip.sty}{%
    \usepackage{parskip}
  }{
    \setlength{\parindent}{0pt}
    \setlength{\parskip}{6pt plus 2pt minus 1pt}}
}{
  \KOMAoptions{parskip=half}}
\makeatother
\usepackage{xcolor}
\ifx\paragraph\undefined\else
  \let\oldparagraph\paragraph
  \renewcommand{\paragraph}[1]{\oldparagraph{#1}\mbox{}}
\fi
\ifx\subparagraph\undefined\else
  \let\oldsubparagraph\subparagraph
  \renewcommand{\subparagraph}[1]{\oldsubparagraph{#1}\mbox{}}
\fi

\providecommand{\tightlist}{%
  \setlength{\itemsep}{0pt}\setlength{\parskip}{0pt}}\usepackage{longtable,booktabs,array}
\usepackage{calc} 
\usepackage{etoolbox}
\makeatletter
\patchcmd\longtable{\par}{\if@noskipsec\mbox{}\fi\par}{}{}
\makeatother
\IfFileExists{footnotehyper.sty}{\usepackage{footnotehyper}}{\usepackage{footnote}}
\makesavenoteenv{longtable}
\usepackage{graphicx}
\makeatletter
\def\maxwidth{\ifdim\Gin@nat@width>\linewidth\linewidth\else\Gin@nat@width\fi}
\def\maxheight{\ifdim\Gin@nat@height>\textheight\textheight\else\Gin@nat@height\fi}
\makeatother
\setkeys{Gin}{width=\maxwidth,height=\maxheight,keepaspectratio}

\makeatletter
\def\fps@figure{htbp}
\makeatother

\makeatletter
\@ifpackageloaded{caption}{}{\usepackage{caption}}
\AtBeginDocument{%
\ifdefined\contentsname
  \renewcommand*\contentsname{Table of contents}
\else
  \newcommand\contentsname{Table of contents}
\fi
\ifdefined\listfigurename
  \renewcommand*\listfigurename{List of Figures}
\else
  \newcommand\listfigurename{List of Figures}
\fi
\ifdefined\listtablename
  \renewcommand*\listtablename{List of Tables}
\else
  \newcommand\listtablename{List of Tables}
\fi
\ifdefined\figurename
  \renewcommand*\figurename{Figure}
\else
  \newcommand\figurename{Figure}
\fi
\ifdefined\tablename
  \renewcommand*\tablename{Table}
\else
  \newcommand\tablename{Table}
\fi
}
\@ifpackageloaded{float}{}{\usepackage{float}}
\floatstyle{ruled}
\@ifundefined{c@chapter}{\newfloat{codelisting}{h}{lop}}{\newfloat{codelisting}{h}{lop}[chapter]}
\floatname{codelisting}{Listing}

\makeatother
\makeatletter
\@ifpackageloaded{caption}{}{\usepackage{caption}}
\@ifpackageloaded{subcaption}{}{\usepackage{subcaption}}
\makeatother
\makeatletter
\@ifpackageloaded{tcolorbox}{}{\usepackage[skins,breakable]{tcolorbox}}
\makeatother
\makeatletter
\@ifundefined{shadecolor}{\definecolor{shadecolor}{rgb}{.97, .97, .97}}
\makeatother
\usepackage{float}
\makeatletter
\let\oldlt\longtable
\let\endoldlt\endlongtable
\def\longtable{\@ifnextchar[\longtable@i \longtable@ii}
\def\longtable@i[#1]{\begin{figure}H]
\onecolumn
\begin{minipage}{0.5\textwidth}
\oldlt[#1]
}
\def\longtable@ii{\begin{figure}H]
\onecolumn
\begin{minipage}{0.5\textwidth}
\oldlt
}
\def\endlongtable{\endoldlt
\end{minipage}
\twocolumn
\end{figure}}
\makeatother
\journal{Cold Regions Science and Technology}
\ifLuaTeX
  \usepackage{selnolig}  
\fi
\usepackage[]{natbib}
\IfFileExists{bookmark.sty}{\usepackage{bookmark}}{\usepackage{hyperref}}
\IfFileExists{xurl.sty}{\usepackage{xurl}}{} 
\hypersetup{
  pdftitle={Retrieving snow depth distribution by downscaling ERA5 Reanalysis with ICESat-2 laser altimetry
},
  pdfauthor={Zhihao Liu; Simon Filhol; Désirée Treichler},
  pdfkeywords={Snow depth, Subgrid variability, Laser
altimetry, Statistical downscaling, ICESat-2, XGBoost},
  colorlinks=true,
  linkcolor={blue},
  filecolor={Maroon},
  citecolor={Blue},
  urlcolor={Blue},
  pdfcreator={LaTeX via pandoc}}

\DeclareUnicodeCharacter{2032}{'}
\DeclareUnicodeCharacter{200B}{}
\begin{document}



\begin{frontmatter}
\title{Retrieving snow depth distribution by downscaling ERA5 Reanalysis with ICESat-2 laser altimetry
}
\author[1]{Zhihao Liu%
\corref{cor1}%
}
 \ead{zhihaol@geo.uio.com} 
\author[1,2]{Simon Filhol%
}
 \ead{simon.filhol@geo.uio.no} 
\author[1]{Désirée Treichler%
}
 \ead{desiree.treichler@geo.uio.no} 
 
\affiliation[1]{organization={Department of
Geosciences, University of Oslo},city={Oslo},country={Norway}}
\affiliation[2]{organization={Météo-France - CNRS, CNRM UMR 3589, Centre d'Etude de la Neige}, city={Grenoble},country={France}}
\cortext[cor1]{Corresponding author}

\begin{abstract}
Estimating the variability of snow depth in remote areas poses significant challenges due to limited spatial and temporal data availability. This study uses snow depth measurements from the ICESat-2 satellite laser altimeter, which are sparse in both space and time, and incorporates them with climate reanalysis data into a downscaling-calibration scheme to produce monthly gridded snow depth maps at microscale (10~m). Snow surface elevation measurements from ICESat-2 along profiles are compared to a digital elevation model to determine snow depth at each point. To efficiently turn sparse measurements into snow depth maps, a regression model is fitted to establish a relationship between the retrieved snow depth and the corresponding ERA5 Land snow depth. This relationship, referred to as subgrid variability, is then applied to downscale the monthly ERA5 Land snow depth data. The method can provide timeseries of monthly snow depth maps for the entire ERA5 time range (since 1950). We observe that the generic output should be calibrated by a small number of localized control points from a one-time field survey to reproduce the full snow depth patterns. Results show that snow depth prediction achieved a \(R2\) model fit value of 0.81 (post-calibration) at an intermediate scale (100 m x 500 m) using datasets from airborne laser scanning (ALS) in the Hardangervidda region of southern Norway, with still good results at microscale (\(R2\) 0.34, RMSE 1.28m, post-calibration). Bias is greatest for extremes, with very deep/shallow snow depths being under- and overestimated, respectively. Snow depth time series modeled at the site level have a slightly smaller RMSE than ERA5 Land data, but are still consistently biased compared to measurements from meteorological stations. Despite such localised bias and a tendency towards average snow depths the model reproduces the relative snow distribution pattern very accurately, both for peak snow (Spearman's {$\rho$} 0.77) and patchy snow meltout in late spring (Matthews correlation coefficient 0.35). The method relies on globally available data and is applicable to other snow regions above the treeline. Though requiring area-specific calibration, our approach has the potential to provide snow depth maps in areas where no such data exist and can be used to extrapolate existing snow surveys in time and over larger areas. With this, it can offer valuable input data for hydrological, ecological or permafrost modeling tasks.
\end{abstract}

\end{frontmatter}
    \ifdefined\Shaded\renewenvironment{Shaded}{\begin{tcolorbox}[sharp corners, boxrule=0pt, enhanced, frame hidden, breakable, interior hidden, borderline west={3pt}{0pt}{shadecolor}]}{\end{tcolorbox}}\fi


\hypertarget{introduction}{%
\section{Introduction}\label{introduction}}

In a warming world, understanding the spatio-temporal variations of seasonal snow is increasingly important for climate impact assessments, meltwater supply \citep{immerzeel.etal_2020, livneh.badger_2020}, permafrost modeling  \citep{gisnas.etal_2016} and ecological responses \citep{callaghan.etal_2011}. 
Seasonal snow accumulates and melts away once a year in a number of climatic zones from, for instance, the forested regions of the taiga (largest terrestrial ecosystem), the open tundra of the Arctic and many high mountain ranges throughout the world such as the Alps, the Himalayas the Andies \citep{sturm.liston_2021}. 
Being white, snow is easily observable from space. However, its thin nature (typically less than 2 m) makes it challenging to measure snow thickness or mass from space at a large scale and with high repeatability. The complex processes driving snow metamorphism and precipitation estimation are hurdles too to rely on models for assessing seasonal snow dynamics. Observing and modeling seasonal snow at a large scale therefore remains a challenge \citep{tsang.etal_2022, mudryk.etal_2020}. This is exacerbated in remote and complex terrain with limited data availability \citep{bormann.etal_2018}.

A major persistent gap is observing or estimating snow depth and water equivalent over mountain regions. In-situ point-based snow stations are generally located in gentle terrain at lower or mid-elevation, often under-representing rugged and higher elevation  \citep{fassnacht.etal_2018}. Employing sensors on airplanes or UAVs (unmanned aircraft vehicles) offers very high-resolution data \citep{buhler.etal_2016, PAINTER2016139, parr2020}, but are costly, of limited spatial extent and potentially limited by legal flight restrictions.  However this can be an effective approach for watershed-scale snow depth mapping \citep{deems.etal_2013}.

Otherwise, space-borne technologies can cover systematically large to global extents at regular intervals. Two main technologies exist to retrieve snow depth: radar and laser altimetry. Radar has several issues with footprint size, penetration depth, snow wetness, and modeling effects due to scattering \citep{dietz.etal_2012}. For instance, passive microwave radar with a footprint size of up to 25 km is insufficient to capture the heterogeneous snow depth in mountain terrain \citep{tsang.etal_2022, mudryk.etal_2020}. 
Despite recent technology development in C-band radar enabling snow depth retrieval at resolutions ranging from 500 meters to 1 kilometer under dry snow conditions \citep{lievens.etal_2019, lievens.etal_2022},  finer-scale snow depth retrieval remains poorly addressed (e.g. 100 m resolution, \citealt{grunewald.etal_2010, mott.etal_2018}). 
Another approach for high-resolution snow depth retrieval is to use laser scanning or photogrammetry. Both methods provide snow depth by collecting surface height data pre- and post-snowfall, enabling snow depth mapping by differencing elevations. Recently, there has been a significant effort to retrieve snow depth by combining snow-free digital elevation models (DEMs) with space-borne lidar measurements of the snow surface from ICESat \citep{treichler.kaab_2017} and its successor ICESat-2 \citep{neuenschwander.pitts_2019}. This cutting-edge satellite offers high-resolution, accurate lidar elevation profile measurements of the Earth's surface, including snow-covered terrains.

\citet{deschamps-berger.etal_2023} derived snow depth from the ICESat-2 ATL06 products and reported an accuracy of 0.2 m (bias) and a precision (normalized median absolute deviation; NMAD) of 0.5 m for low slopes and 1.2 m for steeper areas over the upper Tuolumne basin, California, USA.
\citet{enderlin.etal_2022} found that snow depth estimates based on ICESat-2 data had a median absolute deviation (MAD) ranging from 0.2 m for slopes \textless{} 5° to over 1 m for slopes \textgreater{} 20°.
\citet{besso.etal_2024} questioned the varying accuracy of the ICESat-2 ATL08 product and developed a self-defined processed elevation product, which achieved a MAD of 0.14 m to 0.20 m and root mean square error (RMSE) of 0.18 m to 0.33 m for the Tuolumne Basin and Methow Valley, USA. 
These studies have uncovered ICESat-2 as an emerging and cost-efficient data source for snow depth and also brought attention to challenges associated with the data and elevation differencing workflow, primarily stemming from discrepancies and spatially/temporally varying inconsistencies between reference DEMs and ICESat-2. Therefore, implementing and improving this workflow requires careful co-registration and bias correction on DEMs \& ICESat-2. Additionally, the sparse nature of ICESat-2 measurements presents another significant challenge to deriving comprehensive snow-depth maps: how can we extrapolate both spatially and temporally to areas outside of the ICESat-2 measurement profiles?

Another approach to understanding snow dynamics is through snow modeling. Researchers primarily use two modeling strategies to study these dynamics, namely process-based and statistical approaches. Process-based models \citep{lehning.etal_2006, liston.elder_2006, vionnet2012, kim.etal_2021b} incorporated physical processes, which are driven by meteorological forcing data and yield gridded snow depth products. \citet{mazzolini.etal_2024} combined snow depth transects from the high-resolution ICESat-2 ATL03 product with snow modelling in a data assimilation framework. They spatially propagated sparse ICESat-2 snow profile information using an abstract distance measured in a feature space defined by topographical parameters and snow melt-out climatology. For the 1 km$^2$ Izas catchment in the Spanish Pyrenees, they show that adding snow depth information in addition to the traditionally used fractional snow-covered area observations improves the model skill score by 22\%. 
However, these models are computationally costly and thus often struggle to cover large areas or provide fine resolution. Models are hindered by complex near-surface atmospheric processes and limited data on precipitation and wind fields \citep{freudiger.etal_2017}, leading to a new question: how to quantify the subgrid variability of snow depth \citep{clark.etal_2011}? 
The distribution patterns of snow exhibit a notable repeatability year after year due to their dependence on topography, vegetation, and consistent synoptic weather patterns \citep{sturm.wagner_2010, parr.etal_2020}.
The consistent recurrence of this pattern supports the use of computationally efficient statistical approaches. Many studies seek to establish parameterizations for subgrid variability, such as snow depletion curves, snow depth elevation gradients, snow probability distribution\citep{mendoza.etal_2020}, subgrid snow depth coefficient of variation \citep{liston_2004, he.etal_2019, gisnas.etal_2016} or topographic correlations \citep{helbig.vanherwijnen_2017, mazzolini.etal_2024}. 
Learning and reproducing a snow depth map at fine scales typically involves pattern recognition.
Multiple-linear regression \citep{grunewald.etal_2013, dvornikov.etal_2015}, binary regression trees \citep{revuelto.etal_2014}, random forests \citep{revuelto.etal_2020} or a convolutional neural network \citep{daudt.etal_2023} have been used to predict snow distribution patterns with varied performance (\(R2\) of 0.25-0.91). 
However, these statistical models typically require substantial training data from terrestrial or airborne sensors. Therefore, most models can hardly be transferred to other catchments or seasons \citep{grunewald.etal_2013, revuelto.etal_2020}. 
Another category of statistical models capable of generalizing subgrid variability is commonly known as downscaling models. These models are designed to refine data from coarse, broad-scale grids to localized subgrid levels \citep{maraun_2019}. When applying these models retrospectively or into the future, an important assumption is made: the statistical relationships remain constant over time, a condition known as stationarity. Currently, there is a limited number of statistical downscaling models applied to snow depth \citep{helbig.etal_2024, tryhorn.degaetano_2013}. The primary obstacles involve obtaining sufficient snow depth measurements for training and testing, meteorological forcing data in high resolution and accurately recognising variability through informative features.

In light of these advancements and challenges, we present a method using ICESat-2 data in conjunction with high-resolution DEMs and ERA5-land climate reanalysis data, to effectively generate comprehensive snow depth maps at the hillslope scale. The objectives and workflow of this study are as follows:

\begin{enumerate}
\def\labelenumi{\arabic{enumi}.}
\tightlist
\item Retrieving snow depth from ICESat-2 laser altimetry data across Norway.
\item Using this data to train a machine learning-based downscaling model that accommodates spatial and temporal variations of snow depth in mountain environments.
\item Applying local scaling calibration to downscaled snow depth for a validation area, and validating output snow depth maps at different scales with in-situ observations, gridded snow model products, and meteorological stations.
\item  Discussing the challenges encountered in snow depth retrieval and snow depth downscaling.
\end{enumerate}

The study area of Norway/Hardangervidda mountain plateau is chosen based on the availability of validation data. Still, the workflow is designed to be globally applicable where an accurate DEM and a proper calibration dataset are available. 
To retrieve snow depth measurements from ICESat-2 data, we used nationally and globally available DEMs acquired during snow-free conditions.
Subsequently, we statistically downscaled ERA5 Land from its \textasciitilde9 km native resolution into 10 m  using ICESat-2 snow depth measurements and a machine learning algorithm.
To our knowledge, this study marks the first attempt to use ICESat-2 data to downscale ERA5 Land data, and the first attempt to propagate ICESat-2 snow depths in time and space using a statistical approach.

\hypertarget{study-area-and-data-setting}{%
\section{Study Area and Data
Setting}\label{study-area-and-data-setting}}

\begin{figure}[H]
{\centering \includegraphics[width=\textwidth,height=\textheight]{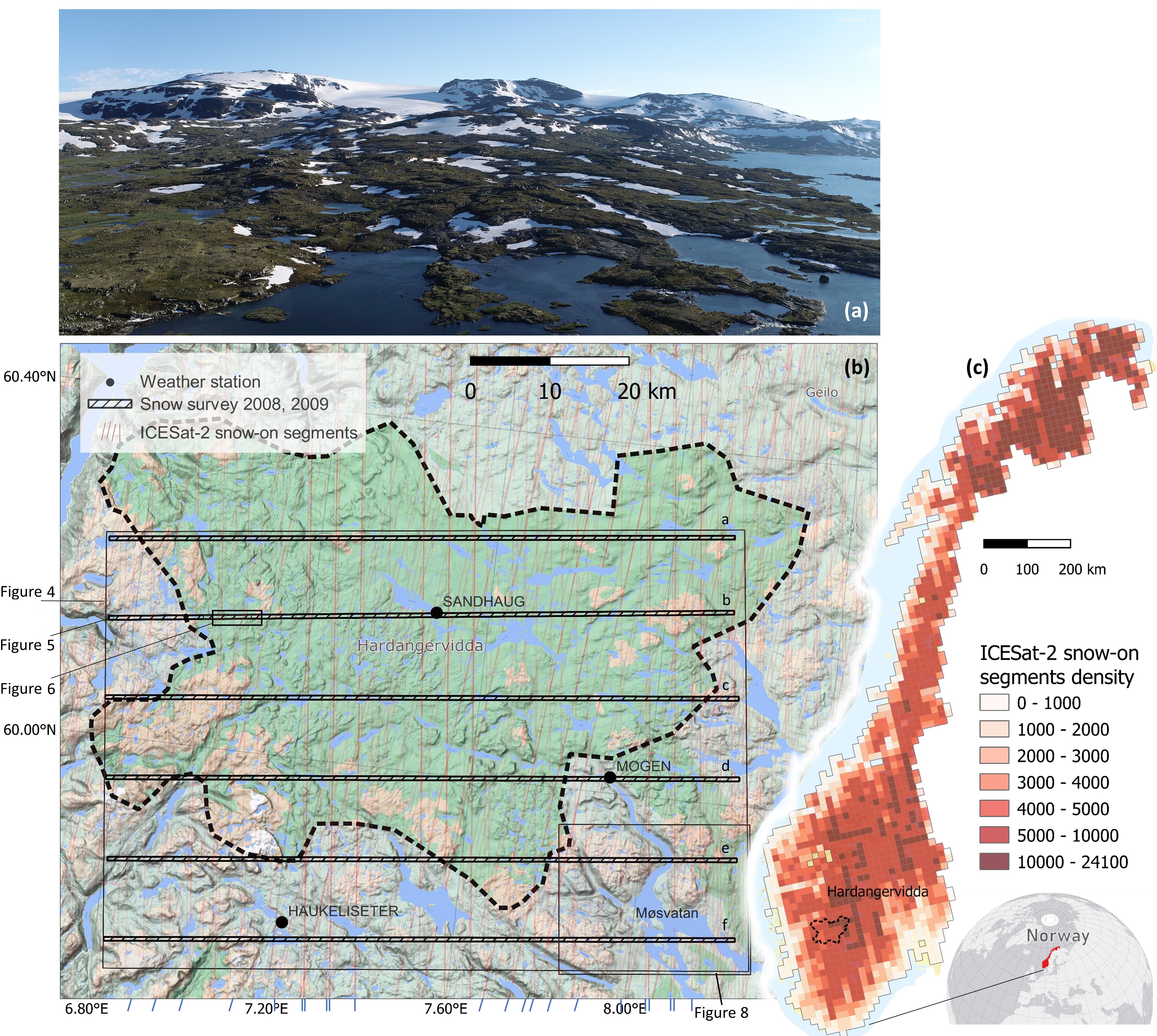}

}

\caption{\label{fig-study_area}Map of the study area, Hardangervidda mountain plateau. A photo of the typical summer landscape is displayed at the top (a), looking south towards the Finse area and Hardangervidda. Photographer: Simon Filhol). The main panel (b) shows ICESat-2 tracks, validation data from the 2008/2009 snow survey and available weather stations. ICESat-2 ATL08 snow-on data from October 2018 to October 2022 are displayed in red. The majority of tracks was surveyed only once and the few repeated tracks appear darker due to higher density. The blue ticks at the bottom highlight the ground tracks acquired during April (peak snow). The right panel shows the location of the validation area and the total number of available snow-on data segments for mainland Norway (c).}

\end{figure}

Norway, located on the western side of the Scandinavian Peninsula in northwestern Europe, spans latitudes from 57° 58′N to 71° 11′ N and longitudes from 4° 40′ E to 30° 58′ E. The country features a diverse topography, ranging from lowland valleys to high mountains (highest peak at 2469 m a.s.l.). In winter, westerly winds bring moisture, resulting in substantial snowfall from the coast to inland areas. This snow acts as a crucial reservoir for hydropower, emphasizing the importance of estimating snow mass in mountain environments. 

Hardangervidda, our validation area, is the largest mountain plateau in northern Europe, approximately 6,500 km². Its comparably flat terrain is nevertheless covered by hills and troughs exposed to high winds and heavy snowfall (Figure~\ref{fig-study_area}). The plateau predominantly lies above 1,000 meters above sea level (m a.s.l.), featuring a low alpine ecosystem with grass heaths, dwarf shrubs, and higher areas with bare rock or lichen marsh tundra. The eastern region is characterized by its open terrain with numerous lakes and streams. The western and southern areas, reach up to 1,700 m a.s.l. and act as significant orographic barriers to the prevailing westerly wind flow. As moist air masses encounter this mountain range, they are lifted and cooled, leading to increased precipitation on the windward slopes and a subsequent decrease of precipitation on the leeward side. Snow accumulation typically begins in mid to late September at higher elevations, peaking around late April. Mean annual precipitation ranges from 750 mm to as much as 3,000 mm over relatively short distances, with approximately 50-60\% of this annual precipitation falling as snow \citep{ketzler.etal_2021}.

\hypertarget{icesat-2-atl08-elevation-data}{%
\subsection{ICESat-2 ATL08 elevation data}\label{icesat-2-atl08-elevation-data}}

Launched in September 2018, ICESat-2 is equipped with the Advanced Topographic Laser Altimeter System (ATLAS), which provides photon-counting lidar measurements at a global scale \citep{neuenschwander.pitts_2019}. ATLAS comprises three parallel beam pairs separated by 3.3 km on the ground. Each beam pair includes a strong and weak beam separated by 90 m. The ATLAS system emits a pulse every 0.7 m along the track, covering a circular footprint with a diameter of \textasciitilde15 m.  At mid-latitudes, ICESat-2 ground tracks are generally not repeated but shifted for each overpass to maximise global spatial coverage. 
The ICESat-2 ATL08 product \citep[level L3A, version 5,][]{neuenschwander.etal_2021a} offers elevation data in fixed segment sizes of 100 meters along the ground track \citep{neuenschwander.etal_2022}. 
For each segment, there are five geolocations (subsegments) in 20 m intervals. Instead of using the mean elevation of the 100 m segment \citep[\emph{h\_te\_mean}, e.g.,][]{enderlin.etal_2022}, we used the subsegment height referring to best-fit terrain elevation at the midpoint location of the segment (\emph{h\_te\_best\_fit\_20m\_2}). 
The sub-segment heights are determined through polynomial fitting to terrain photons with slope correction and weighting \citep{neuenschwander.etal_2022}. 
Norway has a total of 3,968 ATL08 data granules available for analysis from 14 October 2018, to 12 October 2022. After removing invalid data, and excluding permanent ice and inland water, our dataset consists of 13,197,376 segments, including 4,778,904 snow-free segments on land and 8,418,472 segments with snow cover over land. The coverage of snow-on segments is displayed in Figure~\ref{fig-study_area} with blue ticks representing the ground tracks from April of one of the four years (as ground tracks are not repeated).

\hypertarget{snow-off-elevation-data}{%
\subsection{Snow-off elevation data}\label{snow-off-elevation-data}}

As reference ground for snow depth retrieval, we employ the Norwegian~DTM1~elevation model \citep[DTM1,][]{kartverket_2022}, a 1 m lidar-based product acquired by Kartverket between 2016 and 2022. As a sensitivity test of DEM resolution, we also use the 10-meter resolution variant from the same data provider, DTM10 \citep{kartverket_2022}. To demonstrate the workflow's applicability in areas without lidar-based elevation products, we incorporate global DEMs such as Copernicus GLO-30 \citep{europeanspaceagency_2021}, hereafter referred to as COP30, and FABDEM \citep[][Forest And Buildings Removed Copernicus DEM, hereafter referred to as FAB]{hawker.etal_2022} as reference ground. COP30  is a 30-meter-resolution Digital Surface Model (DSM) acquired between December 2010 and January 2015 through synthetic aperture radar interferometry (TanDEM-X mission). FAB, a variant of COP30, eliminates buildings and trees using the random forest algorithm, enhancing accuracy. FAB serves as a reference for comparison with COP30.

\hypertarget{large-scale-reanalysis-data}{%
\subsection{Large-scale reanalysis data}\label{large-scale-reanalysis-data}}

ERA5 Land hourly data (version 5) \citep{munozsabater_2021a} is an ECMWF (European Centre for Medium-Range Weather Forecasts) reanalysis product covering the period from 1950 to the present. It describes water and energy cycles over global land areas with over 50 variables at a spatial resolution of
approximately 9 km \citep{munoz-sabater.etal_2021}. This reanalysis data supplies the necessary forcing data
for the downscaling model to generate sub-grid products while also accounting for input errors in the model \citep{gunther.etal_2019, pflug.etal_2021}. ERA5 Land's snow depth data (\emph{sde}) represents the instantaneous snow thickness on the ground for the elevation of each grid cell, excluding snow on vegetation canopy \citep{munozsabater_2021a}. Additionally, the ERA5 Land monthly \citep{munozsabater_2021} dataset contributes instantaneous wind fields (\emph{u10, v10}) at 10 m above the land surface.

\hypertarget{validation-data}{%
\subsection{Validation data}\label{validation-data}}

Our validation methods include ALS surveys, seNorge snow model data, Sentinel-2 satellite imagery, and meteorological station data. These diverse sources offer both spatial and temporal contexts for evaluating model performance:

\begin{itemize}
\tightlist
\item   The ALS survey: The survey by \citet{melvold.skaugen_2013} provides 2 m gridded snow depths data for two winters over Hardangervidda. The survey encompassed six flight lines apart in 10 km intervals, each extending 80 km in an east-west direction with a crossline scanning width of 500 m (Figure~\ref{fig-study_area}). The data were collected between 3-21 April 2008, 21-24 April 2009, and 21 September 2008 (snow-free reference). During the autumn collection period, the ground was in nearly bare condition except for few perennial snow patches\citep{melvold.skaugen_2013}. The snow depth maps were regridded (averaged) to 10 m spatial resolution for this study.  Figures \ref{fig-snow_micro_2008} and \ref{fig-meso_chart}show snow depth data for April 2008 for parts of flight line b (indicated in Fig.~\ref{fig-study_area}).
\item  The seNorge data (\href{http://www.senorge.no/}{www.senorge.no}) employs a snow model that predicts snow depth based on interpolated precipitation and temperature station observations (seNorge2018 v23.09) \citep{saloranta_2012,saloranta_2016}. It offers daily snow depth maps at a 1 km × 1 km grid resolution and is available via the public archive service Thredds\footnote{MET Norway's Thredds API: \url{https://thredds.met.no}. Last access: 11 Sep 2023.} of the Norwegian Meteorological Institute (MET Norway). We aggregated the daily snow depth from seNorge into monthly average values.
\item  Meteorological stations: this study compared the snow depth time series with three available meteorological stations in the region (Figure~\ref{fig-study_area}). The weather station Sandhaug is located 50 m north of one of the ALS flight lines at an elevation of 1,250 m above sea level (a.s.l.). The other station Mogen (954 m a.s.l.) is situated directly along one of the flight line. Additionally, Haukeliseter (990 m a.s.l.) is positioned between two flight lines. Monthly mean snow depth data were retrieved from MET Norway's Frost API\footnote{Frost API, MET Norway's
    archive of historical weather and climate data: \url{https://frost.met.no}. Last access: 11 Sep
    2023.}. Due to harsh observing conditions in our validation area, all station observations carry a median confidence level indicated by a quality flag of 2.
\item  Sentinel-2 satellite imagery \citep[L2A;][]{copernicussentinel-2_2022} was used for visual data quality checks and validation of the presence/absence of snow  in the Lake Møsvatn area (Figure~\ref{fig-study_area}).
\end{itemize}

\hypertarget{calibration-data}{%
\subsection{Calibration data}\label{calibration-data}}

For calibration purposes, we identified 7103 points where ICESat-2 tracks overlapped with ALS strips collected in April 2008 (Figure~\ref{fig-study_area}). These points, distributed across the entire study area, were selected as control points to represent the study area while the rest of the area remains unseen to the downscaling model. The outputs from the downscaling model are then calibrated against these selected control points to ensure accuracy. 

\hypertarget{methodology}{%
\section{Methodology}\label{methodology}}

\hypertarget{icesat-2-snow-depth-retrieval}{%
\subsection{ICESat-2 snow depth
retrieval}\label{icesat-2-snow-depth-retrieval}}

\begin{figure}[H]
\centering
{ \includegraphics[width=0.95\textwidth]{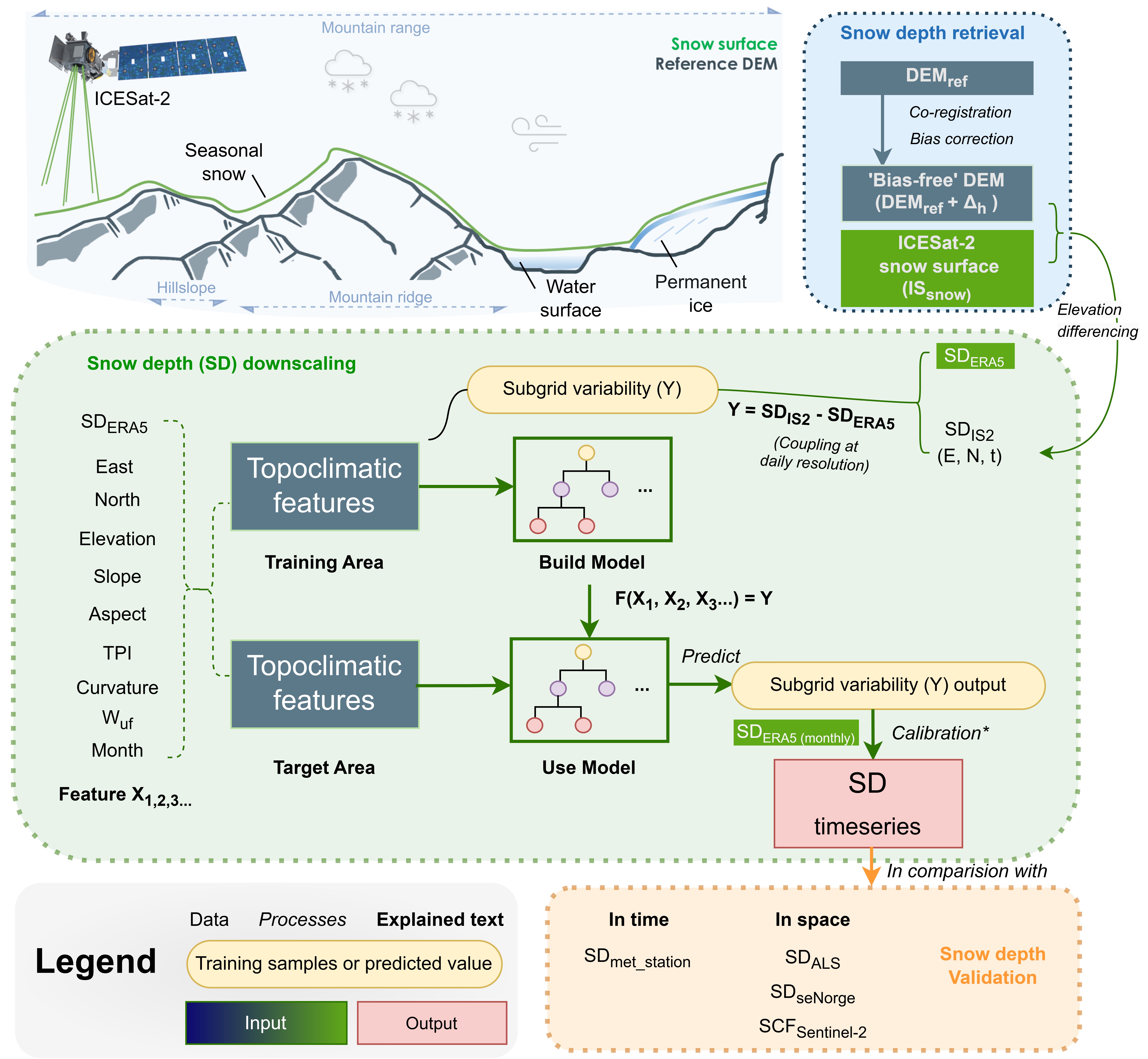}
}
\caption{\label{fig-flow_chart}Flow chart of the snow depth (SD) retrieval
and downscaling-calibration scheme. First,  snow depth is extracted by elevation differencing. This step includes co-registration and bias correction of the DEM (blue box). Consequently, a tree-structure-based regressor for downscaling
is trained and implemented to predict local variability of snow depth in
any location and at any time (green box). The resulting snow depth map time series are validated over time and in space (orange box). For full definitions of acronyms, see Section \ref{methodology}. Satellite graphics source: NASA.gov}
\end{figure}

Snow depth (\(SD_{IS2}\)) was derived from ICESat-2 high-resolution elevation measurements through an elevation differencing workflow. ICESat-2 ATL08 data were categorized into snow-on (\(IS_{snow}\)) and snow-free segments based on attributes and flags present in the ATL08 data: the snow mask (segment\_snowcover) from the National Oceanic and Atmospheric Administration (NOAA) daily snow cover product
  \citep{neuenschwander.etal_2022}, as well as the (brightness\_flag). Snow-free segments were used for DEM co-registration and subsequent bias correction (Figure~\ref{fig-flow_chart}, see \ref{co-registration}, \ref{bias-correction} ). 
The correction involved estimating the discrepancy (\(\hat{\Delta_{h}})\) between the reference DEM (\(DEM_{ref}\)) and ICESat-2 snow-free measurements:
\[
SD_{IS2} = IS_{snow} - (DEM_{ref} + \hat{\Delta_{h}})
\]
Moving terrain such as water surfaces and permanent ice were excluded from the analysis using the segment\_landcover reference mask from the ATL08 product \citep[based on the Copernicus Global Land Cover
dataset at 100-meter spatial resolution,][]
{neuenschwander.etal_2022}.

The snow depth retrieved at each point (Easting, Northing, time) provides spatially and temporally incomplete information on snow dynamics. To overcome this, we obtained daily maximum snow depth from ERA5 Land and interpolated linearly in space for each desired point (E, N, t). The ICESat-2 snow depth is then subtracted from the interpolated ERA5 Land snow depth to create a localized dependent variable, denoted as subgrid variability (\(Y\)) (Figure~\ref{fig-flow_chart}). This variable represents the deviation from the aggregated mean snow depth at each point, used in the subsequent downscaling model (Section~\ref{snow-depth-downscaling}). The downscaling model predicts this subgrid variability, applying it to the ERA5 Land monthly snow depth data to generate snow depth time series. Due to the inherent nature of the downscaling model, raw outputs are biased towards average values while extreme values are underrepresented (\textit{i.e.} conditional bias). Thus, a calibration step (Section ~\ref{calibration}) was added to better represent the full snow depth distribution. Data processing was done in Python with custom scripts that are available in a public GitHub repository (see Section ~\ref{data-availability}). The processing relies on the libraries Xarray \citep{hoyer.hamman_2017} and Pandas \citep{thepandasdevelopmentteam_2024}.  The retrieved snow depth for Norway, spanning from October 2018 to October 2022 is available on Zenodo (DOI: \url{10.5281/zenodo.10048875}).

\hypertarget{co-registration}{%
\subsubsection{Co-registration}\label{co-registration}}

Co-registration is a crucial step to align elevation datasets, with ICESat-2 snow-free data serving as a highly precise and spatially consistent reference. We used a computationally efficient gradient descent-based co-registration algorithm (for details see Supplementary Section A). The process, facilitated by the open-source xDEM tool \citep{xdemcontributors_2021}, was applied to each DEM tile, ensuring accurate alignment across all datasets.

\hypertarget{bias-correction}{%
\subsubsection{Bias correction}\label{bias-correction}}

Bias correction estimates vertical discrepancies (\(\hat{\Delta_{h}})\) between DEMs and ICESat-2 data. Given that DEMs are often patched together from multiple datasets and various sensors, captured in different seasons and at different resolutions, they must be cautiously used as a reference ground surface \citep{hugonnet.etal_2022}.
\citet{magruder.etal_2021} used ICESat-2 elevations to correct DEMs, taking canopy and slope into account. On the other hand, \citet{tian.shan_2021} and \citet{enderlin.etal_2022} found that ICESat-2 ATL08 data underestimates terrain height when compared to the reference DEMs over steep terrain, and thus proposed a slope-dependent bias correction. Our study does not assert which dataset represents ground truth most accurately but focuses on quantifying the discrepancies so that we can exclude it from snow depth elevation differencing.
Essentially, estimating the discrepancies  (\(\hat{\Delta_{h}})\) between two datasets is a regression problem that can be described with physiographic parameters, vegetation conditions, and quality metrics from ICESat-2. 
We employ the XGBoost\footnote{The XGBoost (version 2.0.0) library can be
accessed at \url{https://xgboost.readthedocs.io/}} regression model \citep{chen.guestrin_2016}, a gradient-boosted~decision tree (GBDT) algorithm. The regression model trained based on ICESat-2 snow-free measurements, is later used to predict elevation discrepancies (\(\hat{\Delta_{h}})\) for all other DEM grid cells where ICESat-2 snow-free data are not available (Supplementary Section B). This results in bias-corrected snow depth measurements for all ICESat-2 snow-on data points. The bias correction successfully removed a significant negative skewness in the elevation difference histograms for snow-free data points (Figure S.4). In addition, considerable dependence on first and second-order DEM derivatives (slope and curvature) was detected and removed. After bias correction, NMAD values between ICESat-2 snow-free data points and DEM datasets are reduced from 0.66 m to 0.48 m (DTM1), and from 1.87 m to less than 0.62 m (FAB) (Figure S.4).

\hypertarget{snow-depth-downscaling}{%
\subsection{Snow depth downscaling-calibration}\label{snow-depth-downscaling}}

A second XGBoost regression model is employed to downscale ERA5 Land snow depths. XGBoost has demonstrated its effectiveness in downscaling tasks, such as total water storage anomaly from satellite gravimeter \citep{ali.etal_2023}, precipitation \citep{zhu.etal_2023} or wind speed \citep{hu.etal_2023}. We train the XGBoost regression model on (bias-corrected) ICESat-2 snow depth measurements and use a comprehensive set of topo-climatic features, including snow depth from ERA5 Land (\emph{sde\_era}), east, north, elevation (\emph{h\_te\_best\_fit}), slope, aspect, topographic position index (TPI, \citealt{weiss_2001}; see section \ref{topographic-position-index-tpi}), curvature, planform curvature (\emph{planc}), profile curvature (\emph{profc}), cumulative wind-aspect factor (\(W_{uf}\), see section \ref{wind-aspect-factor}) and month of the year. 
These predictors offer valuable information into the physical driver of snowpack dynamics. For instance, slope and curvatures are basic metrics governing snow accumulation \citep{filhol.sturm_2019}. We compute these terrain attributes using xDEM \citep{xdemcontributors_2021} based on DTM10 with Zevenbergen \& Thorne algorithm \citep{zevenbergen.thorne_1987} at 10 m resolution, consistent with the model's output resolution. The time-varying wind fields are extracted from ERA5 Land monthly data (see section \ref{wind-aspect-factor}). We assume ERA5 land products to be stationary, allowing us to apply the downscaling model trained on snow observations from 2018 to 2022 to other periods.

The XGBoost regression model uses decision trees in parallel structure to capture nonlinear relationships between snow depth subgrid variability and topo-climatic features. During training, the structure and splits of the trees are guided by the goal of minimizing the prediction error on given loss functions.
Two types of loss functions are employed in separate model versions with slightly different purposes: (1) Square error (reg:squareerror) as a loss function to estimate the conditional mean of the target variable, this was the main method used to generate the spatially distributed maps, and (2) Quantile regression (reg:quantileloss) to give probabilistic predictions, such as Q50 (median), Q25 and Q75, which are used for point-based downscaling at weather station locations, to gain insights into the uncertainties of downscaling over time by validating predictions against \textit{in-situ} observation.  
The adoption of quantile regression is inspired by its successful application in probabilistic forecasting \citep{meinshausen_2006, zhang.etal_2018a}, for which no pre-knowledge of the target variable distribution (\textit{e.g.} a normal distribution) is required while being robust to outliers. Quantile regression is therefore an ideal choice for our downscaling task as it captures the full range of possible snow depth values (\textit{e.g.} extreme values) under varying climatic conditions.

\hypertarget{topographic-position-index-tpi}{%
\subsubsection{Topographic position index
(TPI)}\label{topographic-position-index-tpi}}

The TPI is a metric used to assess slope position and classify various landforms. It quantifies the difference between the elevation of a central pixel and the average elevation of its neighboring pixels (3 x 3 pixels). 
A TPI value of zero or near zero indicates a flat or nearly continuous slope. Positive TPI values suggest that the central pixel is significantly higher than the surrounding areas, forming a ridge or hill.
Inversely, negative TPI values indicate that the central pixel is notably lower than its neighboring areas, signifying a valley. TPI has proven effective in predicting snow distribution in alpine environments \citep{revuelto.etal_2014, cristea.etal_2017}. 
To represent landforms at different scales, we used two additional indices: tpi\_9 (calculated in 9 x 9-pixel windows, equivalent to 90 m x 90 m) and tpi\_27 (270 m x 270 m).

\hypertarget{wind-aspect-factor}{%
\subsubsection{Cumulative wind-aspect factor}\label{wind-aspect-factor}}

\begin{figure}[H]

{\centering \includegraphics[width=1\textwidth,height=\textheight]{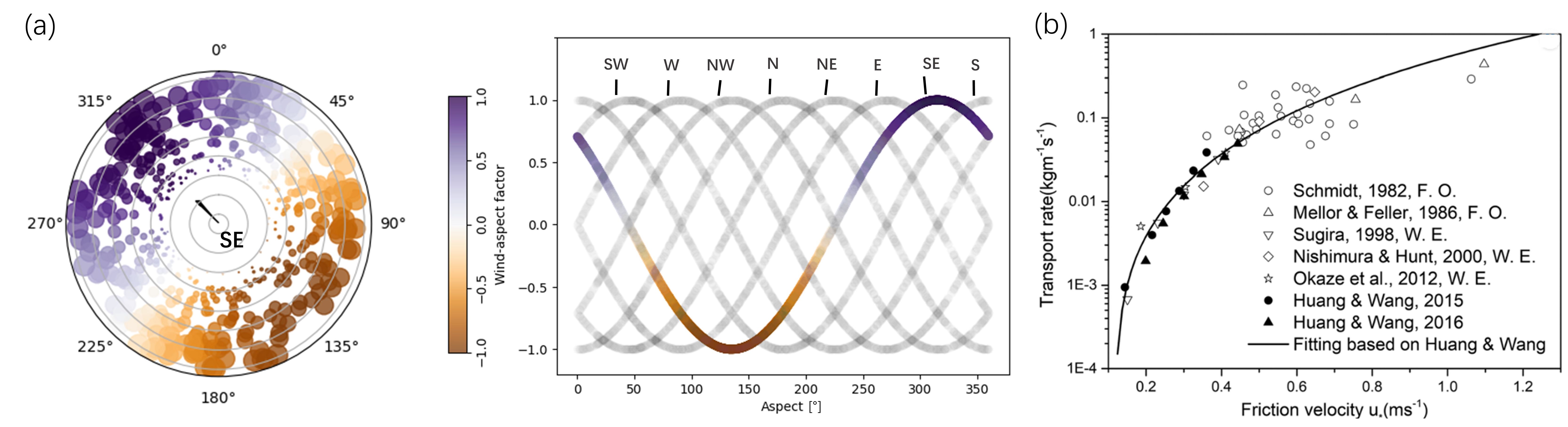}

}

\caption{\label{fig-wf}Quantifying the relationship between wind, aspect
and snow redistribution. (a) The prevailing wind, e.g.~from SE, results in a
negative value on the windward side (snow erosion, $W_{uf_{negative}}$, brown) and positive values on the lee side
(snow deposition, $W_{uf_{positive}}$, blue). Eight cardinal directions are plotted but the function works on any
wind direction. (b) Research suggests that as wind speed
increases, the transport rate increases exponentially with a power of 3 (adapted from \citealt{li.etal_2018}).}

\end{figure}

The wind-aspect factor (\(W_{f}\)) \citep{bennett.etal_2022, dvornikov.etal_2015} serves as a proxy for snow accumulation and erosion on topographic obstacles. 
It assigns positive values on the leeward side and negative values on the windward side of these features. We formulated the relationship between wind and aspect by a cosine function that ranges from -1 to 1 for any prevailing direction (see Figure~\ref{fig-wf}):

\[W_{f} = -cos(aspect - dir_{wind})\]

where \(dir_{wind}\) is the direction of the wind origin with northerly wind (blowing from north to south) referred to as 0°. This study further divided \(W_{f}\) into leeward \(W_{f_{positive}}\) and windward factors \(W_{f_{negative}}\) , multiplied by wind speed to the power of three (Figure~\ref{fig-wf} b) to capture the cumulative effect of wind redistribution for each water year period.

\[W_{uf_{positive}} = \sum W_{f_{positive}}u_{wind}^3\]

\[W_{uf_{negative}} = \sum W_{f_{negative}}u_{wind}^3\]

where \(u_{wind}\) is the monthly average wind speed from ERA5 Land, linearly interpolated to 10 meters resolution. The accumulation begins in September from zero until the next August when the value reaches its maximum. The value does not accumulate when the monthly average snow depth falls below 0.1 m during the annual cycle.

\hypertarget{calibration}{%
\subsubsection{Calibration}\label{calibration}}

In our study, the XGBoost downscaling model is tasked with predicting snow depth under a variety of conditions. We observed that XGBoost tends to produce a conservative estimate close to the mean, likely because the feature set does not consistently explain what critical conditions lead to extreme snow depths. This conservative tendency is a natural outcome of the model's objective to minimize overall prediction error, often resulting in a distribution that skews towards average conditions and under-represents the extremes (known as scaling bias, further discussed in Section \ref{challenges-of-downscaling-snow-depth}). The calibration step scales the modelled snow depth values so that their distribution matches the locally observed snow depth distribution, using quantile mapping \citep{cannon.etal_2015,li.etal_2010}. Thereby, while preserving the relative snow depth predictions of the downscaling model, the calibration removes the scaling bias for each specified quantile by a scaling factor calculated from the control points, ensuring a good representation of local snow depth distribution:
\[ x_{adj} = F_O^{-1}(F_M(x)) \]
\[ \Delta = \frac{x_{adj}}{x} \]
Here, \(x_{adj}\) denotes the calibrated prediction. Determined by the cumulative distribution function (CDF) of the XGBoost downscaling model output \(F_{M}(x)\) and that of the observational control points \(F_{O}(x_{adj})\), the discrepancy between two distribution's quantiles is encapsulated by the scaling factor (\(\Delta\)) for a given ratio.

\hypertarget{model-interpolation-and-assessment}{%
\subsection{Model interpretation and
assessment}\label{model-interpolation-and-assessment}}

To interpret the contribution of topo-climatic features in our tree-based models, we employed the SHAP values \citep{lundberg.etal_2020}, a metric derived from game theory facilitating the understanding of the relative contribution of model predictors (\textit{a.k.a.} features). The SHAP value corresponds to the contributions of each feature to individual predictions. A feature can either play for or against the prediction (positive or negative SHAP value), and the magnitude of its SHAP value shows the significance of the feature role. The sum of all these contributions, plus the base prediction, provides the final prediction.

For fair comparisons,  we aggregated all high-resolution validation datasets to the same resolutions.  Model validation was conducted across two different scales.  At the micro to site scale (respectively 100 to 10 m), we captured snow distribution over typical micro-terrain features such as hills and gullies using the original model output resolution of 10 m. At the mesoscale, we aggregated data into a 100 m x 500 m grid, reflecting the 500 m width of the ALS survey swath, with 100 m intervals in the transect direction to obtain a good characterisation of snow depth variability across different aspects of hills and ridges.  Successful prediction would result in a (near-) perfect data match in probability distribution and ranking correlation. To quantify the downscaling performance, we employ four key statistical metrics. RMSE and \(R2\) scores evaluate the overall accuracy and fitness of the model.  \(R2\) score is computed using the standard implementation of the Python library Scikit-learn: 

\[R2 =  1 - \frac{\mathrm{SS_{res}}}{\mathrm{SS_{total}}} \]

where \(\mathrm{SS_{res}}\) represents the variation in the data that the fitted model does not explain, expressed by the sum of squared residuals between the model output and measured data, and \(\mathrm{SS_{total}}\) is the total variation in the data, i.e., the sum of squared residuals with regard to the mean. The \(R2\) value typically ranges from 1 (perfect fit) to 0, but can be negative if the model is evaluated on different data than used for training (as in our case) and for nonlinear models typically used in machine learning approaches (e.g. XGBoost, as used in this study). A negative \(R2\) value means that the model performs worse than a constant function that always predicts the mean. As R2 is sensitive to the presence of bias, Spearman's rank correlation coefficient (\(\rho\)) is used as a fidelity metric, with a high \(\rho\) indicating good similarity in spatial distribution. The Kolmogorov-Smirnov D statistic (KSD) quantifies the degree of probability distribution matching, with KSD = 0 indicating a perfect match. To compare the patchy snow distribution during melt-out season with binary snow cover data from satellite imagery, we use the area under curve (AUC) and Matthews correlation coefficient (MCC). For these metrics, a value of 1 indicates a perfect match, whereas values of 0.5 (AUC) and 0 (MCC) correspond to random guessing.  

Additionally, we used variograms to quantify the model's ability to capture the spatial heterogeneity of snow depth. The semi-variance (\(\gamma\)) is a measure of spatial variability, calculated for pairs of observations as half the average squared difference between values separated by a specific lag distance (\(l\)) \citep{oliver.webster_2014}:

\[\gamma(l) = \frac{1}{2N(l)} \sum_{i=1}^{N(l)} (z(x_i) - z(x_i + l))^2\]

where \(z(x_i)\) represents the snow depth at location \(x_i\). The variogram indicates the rate at which correlation decreased with distance. By fitting variograms to the sum of the spherical model and Gaussian model for short and long ranges, respectively \citep[following the method of ][]{rolstad.etal_2009, hugonnet.etal_2022}, we identify spatial correlation of snow distribution at different scales. The variograms are computed using the xDEM tool \citep{xdemcontributors_2021}.

\hypertarget{results}{%
\section{Results}\label{results}}

\hypertarget{mesoscale-snow-depth-variability}{%
\subsection{Mesoscale snow depth
variability}\label{mesoscale-snow-depth-variability}}

\begin{figure}[H]
\centering
{\includegraphics[width=\textwidth,height=\textheight]{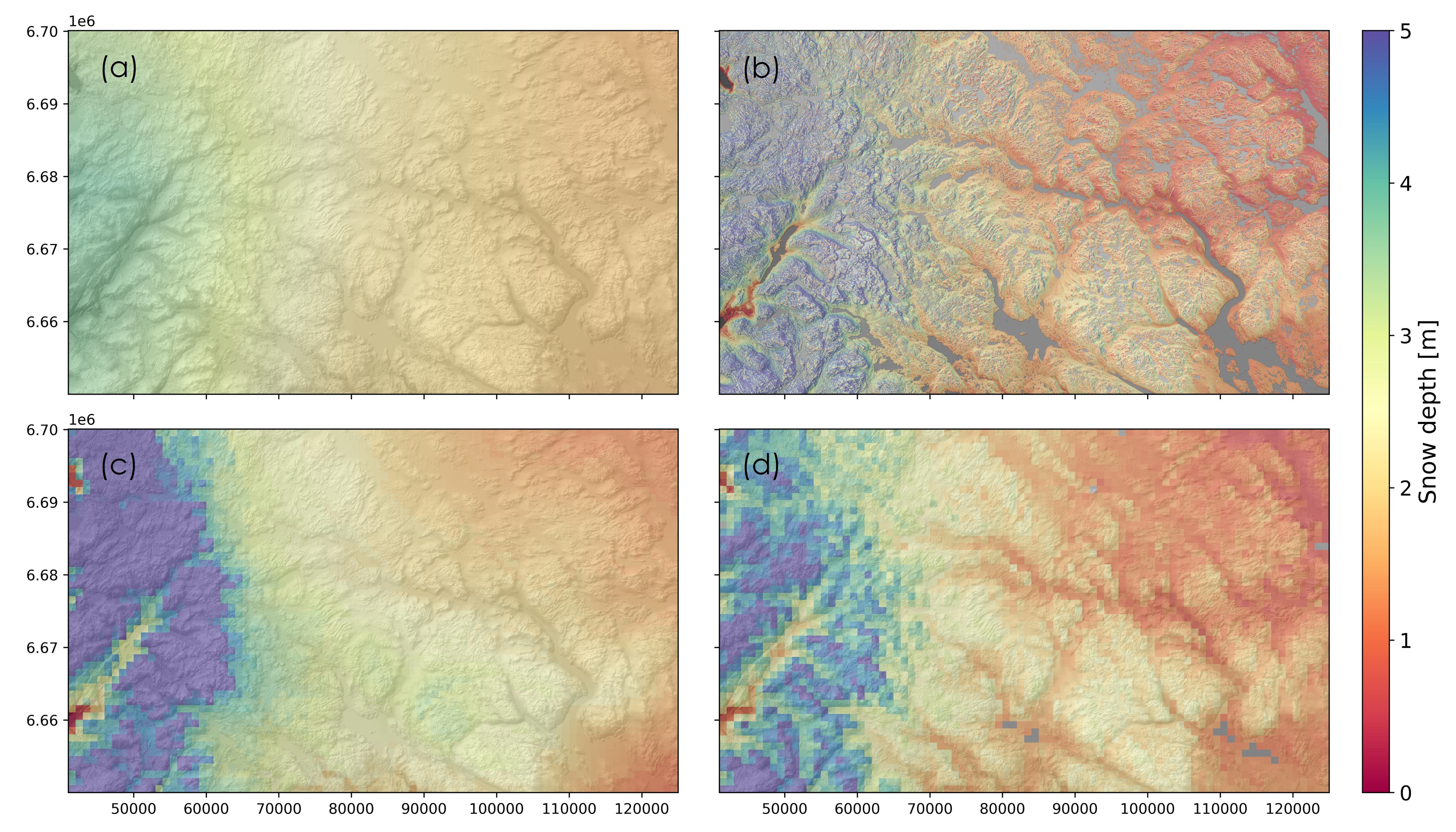}
}
\caption{\label{fig-snow_meso_2008}Spatial distribution of snow depth in the Hardangervidda area, April 2008. (a) Downscaling model input, linearly interpolated snow depth from ERA5 Land. 
(b) Downscaled snow depth output at 10 m resolution. 
(c) seNorge snow depth at 1 km resolution.
(d) Downscaled snow depth aggregated to 1 km resolution. Map coordinates are in meters UTM 33N.}
\end{figure}

\begin{figure}[H]
\centering
{\includegraphics[width=\textwidth,height=\textheight]{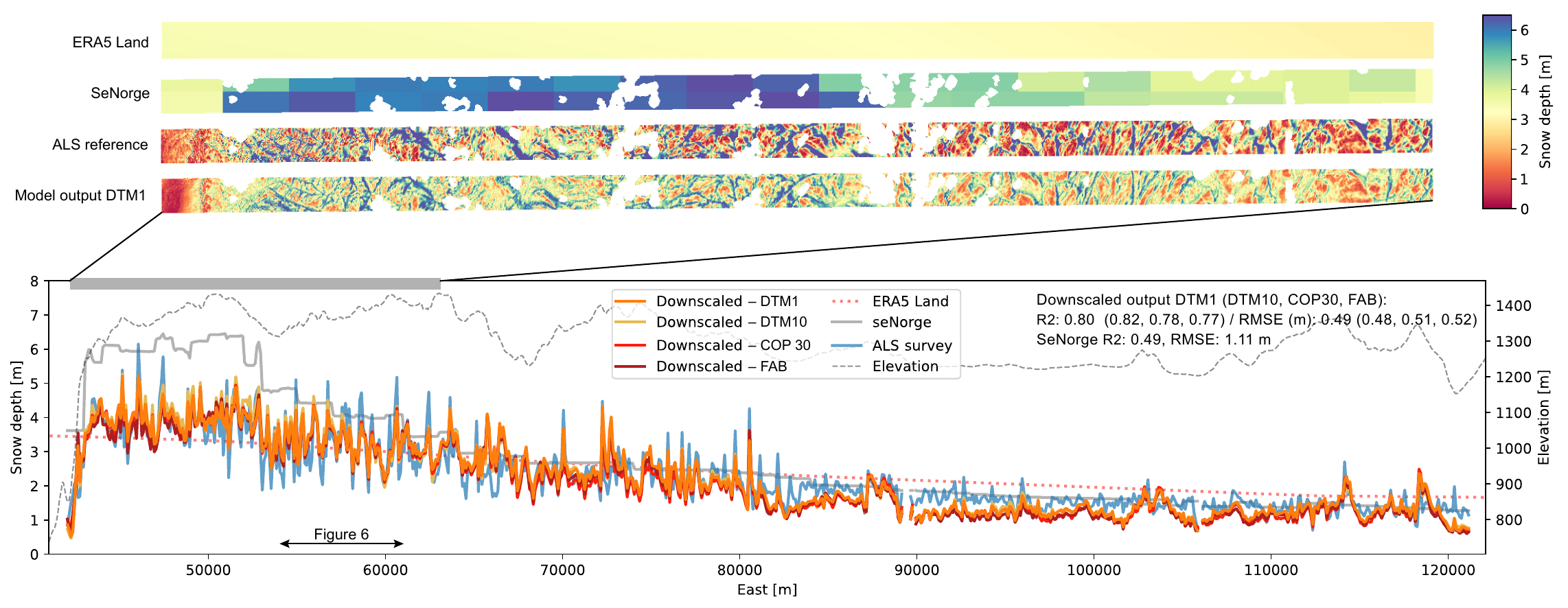}
}
\caption{\label{fig-meso_chart}Snow depth profile of flight line b across Hardangervidda as shown in Figure \ref{fig-study_area}). Data is shown for April 2008 at mesoscale, i.e. averaged to 100 m x 500 m (flight strip width) cells along the flight strip length of 80 km. Model outputs of the different DEMs are nearly identical. The map strips illustrate snow depth from ERA5 Land, seNorge, ALS snow survey and downscaled model output using the DTM1 for the marked section of the datasets, in their native resolution before aggregation. }
\end{figure}

\begin{figure}[H]
\centering
{\includegraphics[width=\textwidth,height=\textheight]{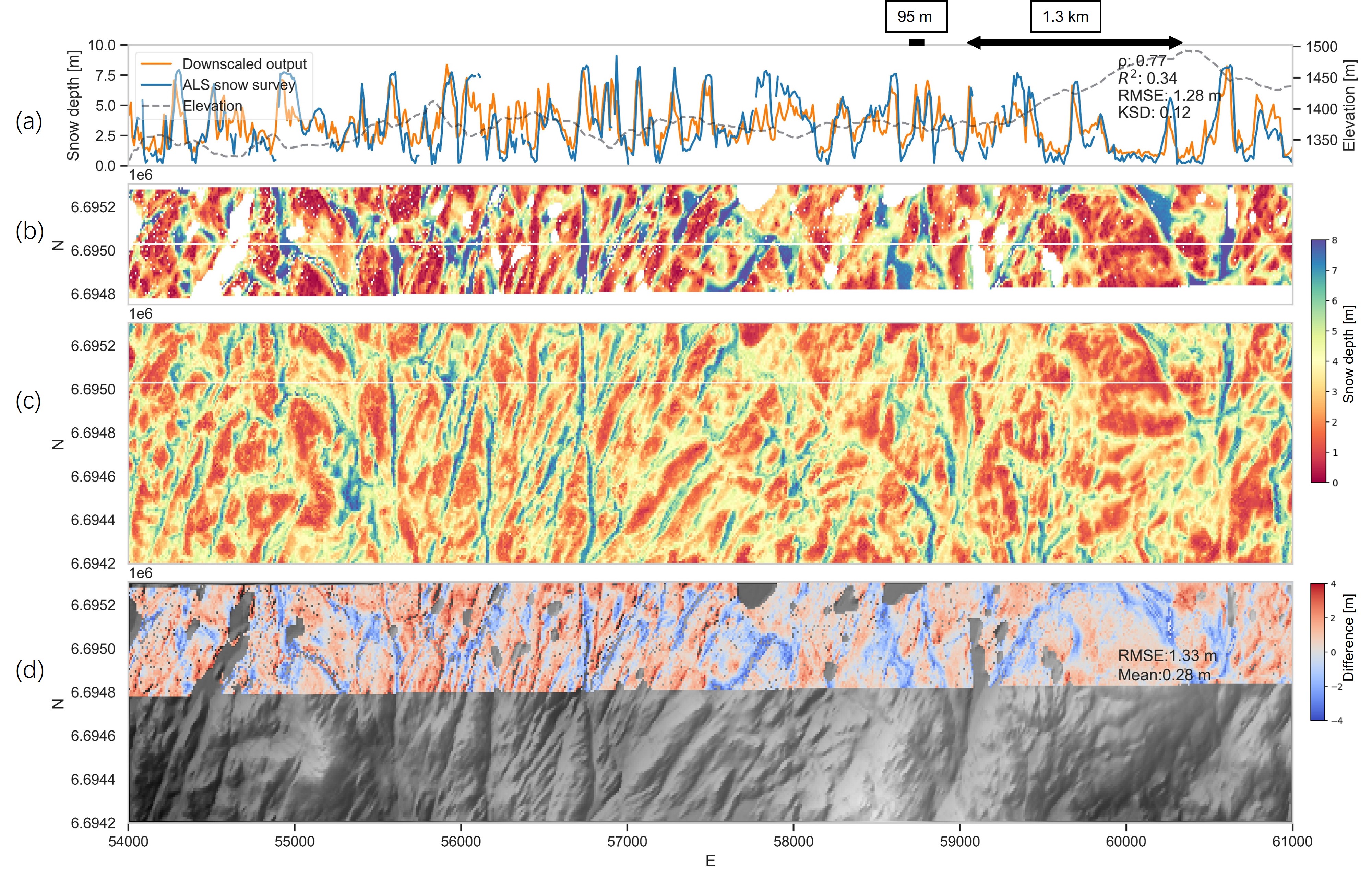}
}
\caption{\label{fig-snow_micro_2008}Microscale (10 m) snow depth comparison in
the west Hardangervidda, April 2008. The validated area is located along flight line b, as shown in the study area map in Figure \ref{fig-study_area} and mesoscale snow depth profile in Figure \ref{fig-meso_chart}. (a) Transect line, marked white in panels b/c, (b) ALS snow survey validation strip with significant snow depth variations, (c)
downscaled snow depth, (d) differences between the ALS data and model output (background: DTM1 DEM). Common horizontal scales of landforms (95m and 1.3 km) are symbolized by black arrows in panel a.}
\end{figure}

Figure~\ref{fig-snow_meso_2008} shows the snow depth maps for April 2008. Model input data, i.e., linearly interpolated snow depth from ERA5 Land (a), only represents large-scale variability. After downscaling, the fine-scale snow depth variability aligns with the topography both at the microscale (10 m, b) and aggregated to 1 km (d). In comparison, seNorge data (c) at the same spatial resolution appear smoother with less spatial variability and overestimate snow depth in the western mountains. The differences between the two data sets are smaller for April 2009 where there was generally less snow in Southern Norway (see Supplementary Section E).  A comparison of April 2008 snow depths at mesoscale for flight line b is shown as a transect in Figure~\ref{fig-meso_chart}. The ALS snow survey, downscaled output (after calibration), seNorge, ERA5 Land data and elevation are aggregated (averaged) to a resolution of 100 m x 500 m, owing the ALS survey's 500 m transect width. All datasets follow a decreasing trend from snow depths exceeding 6 m to the West (close to the coast) to around 2 m to the East (far from the coast). 

The downscaled snow depths reveal impressive details in snow depth spatial variability at the mesoscale that corresponds well to the observed snow depth magnitude and spatial variability captured by the ALS snow survey. The goodness of fit is similar for the other flight strips (Figures S.7, S.8). Across the six flight lines, the calibrated downscaling models score an \(R2\) of 0.81 and an RMSE of 0.53--0.57 m (Table \ref{tab:mesoscale-table}). We achieve nearly as good performance at mesoscale when using the global DEMs COP30 and FAB for ICESat-2 snow depth sample retrieval rather than the Norwegian high-resolution datasets DTM1 and DTM10 (Figure S.7). There are only minimal differences between output snow depth maps and we did not detect any systematic biases, no matter which DEM was used for snow depth sample retrieval. The results are similarly good for the 2009 ALS snow survey (Figure S.8 ), with slightly lower RMSE values of 0.50 m to 0.55 m that can be attributed to generally lower snow depths in 2009.

In comparison, \(R2\) values of the (coarser-resolution) seNorge model data compared to the ALS survey are lower (0.49 and 0.65) and RMSE values higher (1.12 m and 0.60 m). The seNorge model shows little spatial variability, is less sensitive to topographic features and orographic barriers, and  shows about 40 \% overestimation on the western mountain ridges but matches ALS data in the East. \citet{melvold.skaugen_2013} attributed the overestimation of the seNorge model to its strong reliance on weather stations located in low-elevation areas not representative for high mountains. 

Upon closer inspection, ALS ground truth data exhibit many spikes on east-facing slopes (Figure~\ref{fig-meso_chart}), likely resulting from wind redistribution and gravity processes. Our downscaling model acknowledges the importance of the aspect and wind-driven snow accumulation factors (see also Figure~\ref{fig-shap} c,g) but does not always reproduce corresponding snow distribution in the correct places. ERA5 land snow depth input data exhibit a north-south systematic bias, shifting from overestimation to underestimation relative to the ALS data (Figure~\ref{fig-shap} d). The model over-corrects snow depths in the Eastern part of flight line b, resulting in snow depth underestimation compared to the ALS survey. 

\begin{table}[H]
\caption{Statistical comparison of snow depth estimates from different methods (downscaled output with snow retrievals based on different DEMs, seNorge, and ERA5 Land) against ALS snow survey data for all six flight lines (a–f) combined, for the years 2008 and 2009, respectively. \(R2\), KSD and Spearmans' \(\rho\) are unit-less, RMSE is in m.}
\label{tab:mesoscale-table}
\begin{tabular}{@{}lcccc|cccc@{}}
\toprule
    & \multicolumn{4}{c}{April 2008} & \multicolumn{4}{c}{April 2009}  \\ \midrule
Dataset & \(R2\)  & KSD &    \(\rho\) & RMSE & \(R2\)  & KSD &    \(\rho\) & RMSE \\
    DTM1 &   0.81 &   0.09 &   0.88 &   0.53   &   0.78 &   0.07 &   0.87 &   0.50  \\
    DTM10 &   0.82 &   0.12 &   0.88 &   0.56  &   0.79 &   0.08 &   0.88 &   0.513  \\
    COP30 &   0.80 &   0.10 &   0.88 &   0.57 &   0.75 &   0.07 &   0.84 &   0.554  \\
    FAB &   0.81 &   0.11 &   0.89 &   0.56 &   0.77 &   0.07 &   0.86 &   0.54  \\
    SeNorge &   0.49 &   0.21 &   0.84 &   1.12 &   0.65 &   0.13 &   0.86 &   0.60  \\
    ERA5 Land &   -0.79 &   0.27 &   0.71 &   0.82 &   -0.34 &   0.24 &   0.72 &   0.68  \\
\bottomrule
\end{tabular}
\end{table}

\hypertarget{microscale-snow-depth-variability}{%
\subsection{Microscale snow depth
variability}\label{microscale-snow-depth-variability}}

Figure~\ref{fig-snow_micro_2008} compares the 2008 ALS snow survey with the downscaled snow depth at a 10-meter resolution. The area shown, located in the western part of flight line 2 (indicated in Figure \ref{fig-study_area}), exhibits distinct microscale landforms, with sheltered depressions hosting thick snow patches (\textgreater{} 8 m) and wind-exposed hilltops featuring thin snow covers (close to 0 m). This area was also shown in \citet[Figure 3]{melvold.skaugen_2013} to address the effect of spatial resolution on snow depth representation. The terrain features align with the spatial lag that exhibits significant autocorrelation of snow depth in the variogram in Figure \ref{fig-variogram} (e), corresponding to ca.~100 m and 1.3 km.

The transect line (marked in white) in the downscaled output across this varied terrain visually captures most of the observed pattern. Extreme deep/shallow snow depths tend to be biased towards average values, but less so than if scaling calibration is not applied (Figure S.6). While Spearman's \(\rho\) of 0.77 indicates a strong rank correlation,  the relatively low \(R2\) value of 0.34 and high RMSE of 1.33 m (panel d) suggests a reduced statistical agreement compared to the mesoscale analyses. The KSD of 0.12 for the transect line and a mean deviation of 0.28 m for the difference map indicate that the scaling calibration of the downscaled snow depths, a single function applied to the entire study area and not the shown sub-region specifically, does not fully reproduce extreme values and results in a slight overestimation of average snow depths in this sub-region. The residual differences appear to be correlated with terrain features and match areas with remaining snow patches in the DTM1, which we detect by comparing the DTM1 with the ALS snow-off data (Figure S.6 e).

\hypertarget{temporal-variability-of-snow-depth}{%
\subsection{Temporal variability of snow
depth}\label{temporal-variability-of-snow-depth}}

\begin{figure}[H]
\centering
{\includegraphics[width=0.8\textwidth]{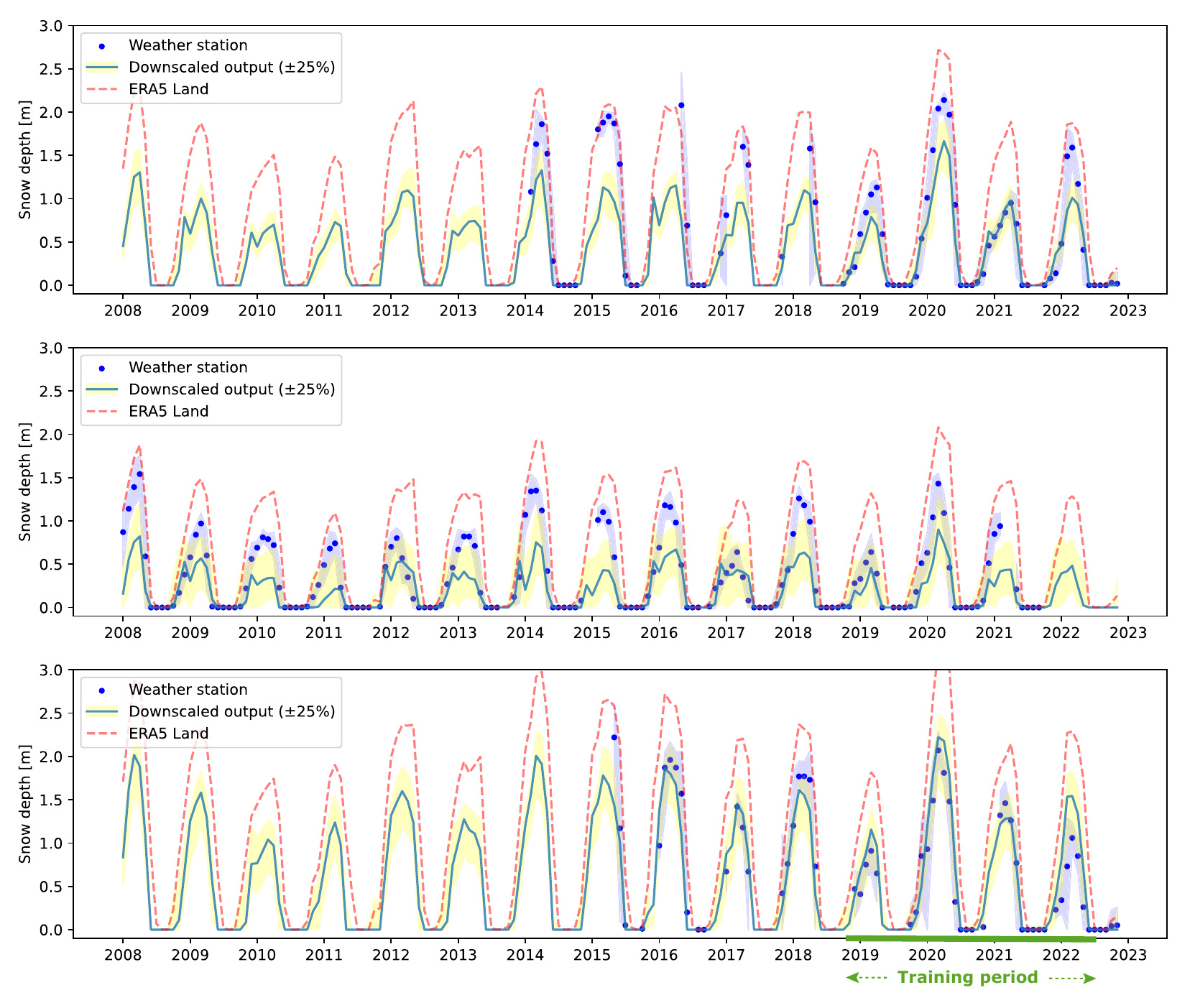}
}
\caption{\label{fig-snow_weatherstation}Time-series validation of snow
depth for three weather stations Sandhaug (a), Mogen (b), and Haukeliseter
(c). Blue Dot: measured monthly mean and minimum/maximum
values (light blue), Blue line: median downscaled snow depth (Q50) and interquartile range (IQR) for 25th--75th
quantiles (yellow). Note: Haukeliseter and
Sandhaug data availability starts from 2015 and 2014, respectively, and
some years have missing data and incorrect measurements.}
\end{figure}

The downscaling model is also able to propagate information in time rather than only space. For this purpose, we used quantile regression for the downscaling model. The uncalibrated model output is shown in Figure~\ref{fig-snow_weatherstation}, to visualize the local bias and its evolution over time. 
Time series data are shown for the data cell (10 m pixel size) corresponding to the locations of the three stations within the Hardangervidda area: Sandhaug (a), Mogen (b), and Haukeliseter (c). The predicted interquartile range Q25--Q75 (IQR, shown in yellow) provides an estimate of the snow depth predictions uncertainty. Q50 (blue line) represents the median prediction. Visually, Q75 shows the best match with the weather station data whereas ERA5 data over- and Q50 model data underestimate measured snow depths (Table \ref{tab:station-table}). 
(\(R2\) of 0.71, 0.60, 0.76) 
The downscaled model (both Q50 and Q75) generally performs similarly to the original ERA5 Land data across all stations. \(R2\) values range from -0.43 (Q50, Mogen) to 0.86 (Q50, Haukeliseter), and do not describe performance well due to significant biases at the individual site level. While the (uncalibrated) model exhibits bias in predicting the absolute snow depths, the high Spearman's \(\rho\) (0.91 to 0.95, ERA5: 0.92 to 0.96) across all three stations suggests that the downscaling successfully maintains the relative ordering of snow depth time series. The better agreement of Q75 with observations (compared to Q50) could reflect either (1) a systematic ~25\% underestimation bias in the model, or (2) the local station measurements corresponding to the Q75 percentile of the model’s predictions for similar conditions.
Bias is primarily observed for high snow depth values during the peak snow season whereas the snow-free season is captured accurately. The peak snow bias is different for the three sites but has a consistent magnitude over time at each site. There are no indications of a better fit for the training period (late 2018 to late 2022) compared to earlier years. This suggests that the bias could be reduced or removed entirely by a scaling calibration such as applied to the maps resulting from the spatial propagation presented above.

\begin{table}[H]
\caption{Statistical comparison of monthly snow depth estimates from different methods (Downscaled output and ERA5 Land) against weather station snow depth. RMSE is in metres.}
\label{tab:station-table}
\begin{tabular}{@{}lcccc|cccc|cccc@{}}
\toprule
        & \multicolumn{4}{c}{Model --- Q50} & \multicolumn{4}{c}{Model --- Q75} & \multicolumn{4}{c}{ERA5 Land}             \\ \midrule
Station   & R2  &  KSD &\(\rho\)&RMSE & R2  &  KSD &\(\rho\)&RMSE & R2  & KSD & \(\rho\)&RMSE\\
Sandhaug     & 0.29  & 0.20 & 0.94 &0.39 & 0.71 &  0.17 &0.95 & 0.29 & 0.73  & 0.21 & 0.96 & 0.44\\
Mogen        & -0.43 &  0.22 & 0.91 & 0.28 & 0.60 &  0.19 &0.91 & 0.23 & 0.47  & 0.31 & 0.92 & 0.46\\
Haukeliseter & 0.86  &  0.10 & 0.94 & 0.26 & 0.76 &  0.23 &0.94 &0.38 & 0.37  & 0.36& 0.94 & 0.79 \\ \bottomrule
\end{tabular}
\end{table}

\hypertarget{validating-snow-occurrence}{%
\subsection{Validating snow occurrence}\label{validating-snow-occurrence}}

\begin{figure}[H]
\centering
{\includegraphics[width=\textwidth,height=\textheight]{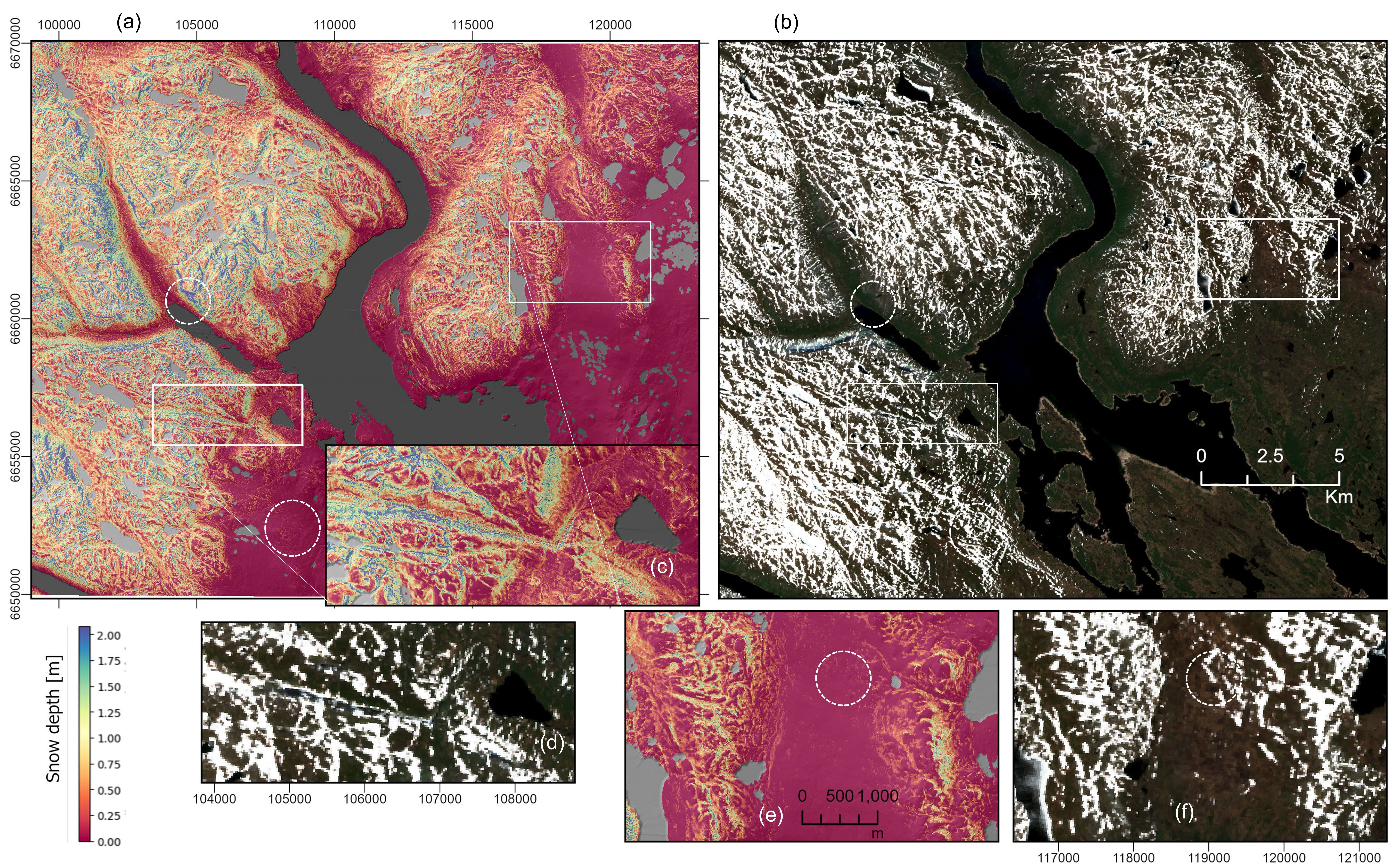}
}
\caption{\label{fig-snow_presence_2020}Validation of snow occurrence
using Sentinel-2 imagery for the Lake Møsvatn area (annotated in Figure \ref{fig-study_area}). (a) modelled snow cover
distribution for June 2020, (b)  Sentinel-2 image on June 24th,
2020. Panels (c) and (d) show local discrepancies in snow distribution, (e) and (f) show correctly modelled snow retention on east-facing versus west-facing slopes.
The overestimation on steep slopes and underestimation in flatter
terrain are annotated with white circles.}
\end{figure}

\begin{figure}
{\centering \includegraphics[width=1\textwidth]{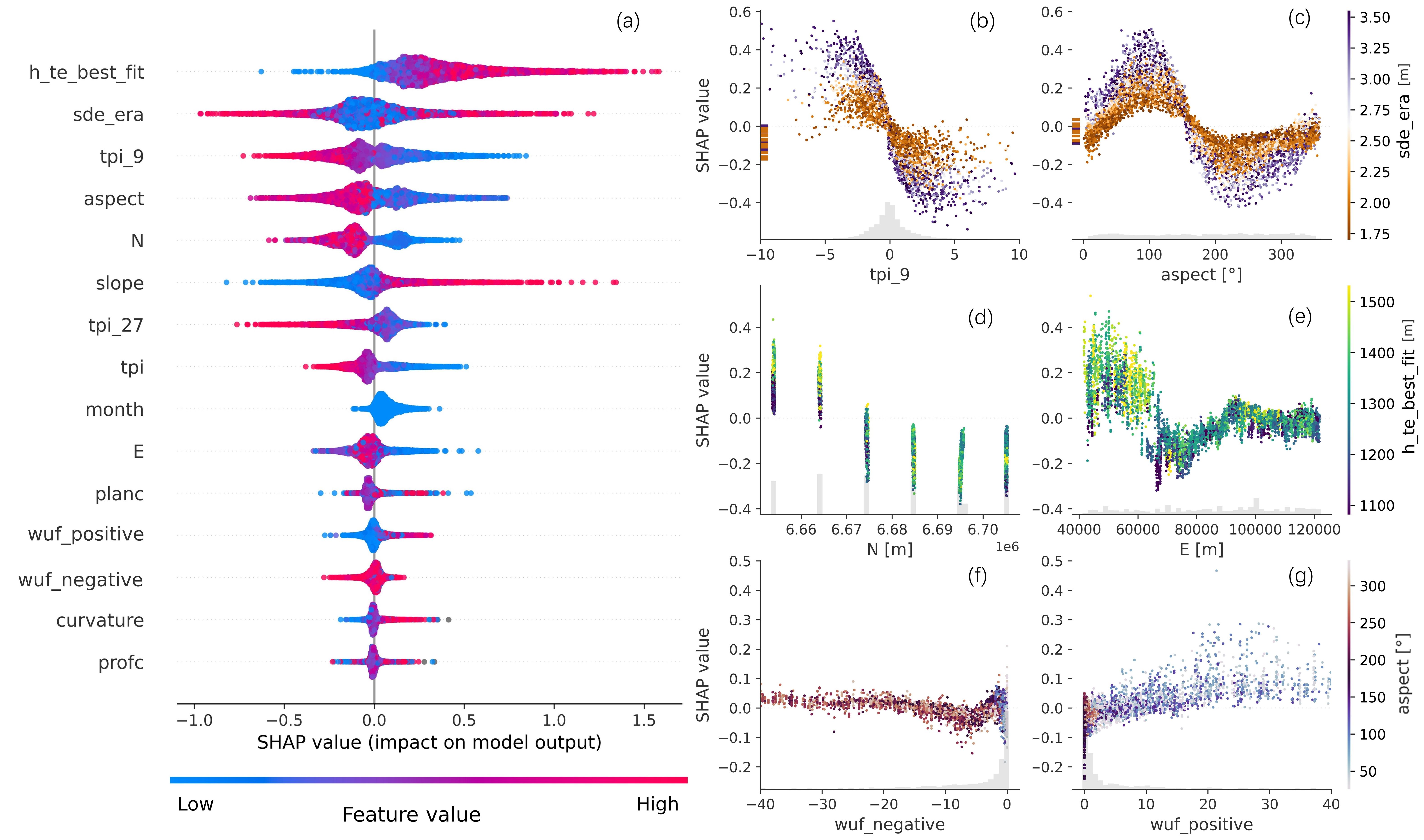}
}
\caption{\label{fig-shap}SHAP interpretation for the downscaling model.
Panel (a) ranks features by their impact on snow depth
prediction. Subsequent panels (b-g) depict SHAP dependence plots,
illustrating how each feature influences model predictions for the ALS survey areas in the Hardangervidda area. A high SHAP value positively influences predicted snow depths, and a low SHAP value causes lower snow depth predictions. Each dot represents an individual 10 m grid cell,  and the light grey histograms on the x-axis show the relative distributions of feature values.}
\end{figure}

\begin{figure}
\centering
{\includegraphics[width=0.6\textwidth]{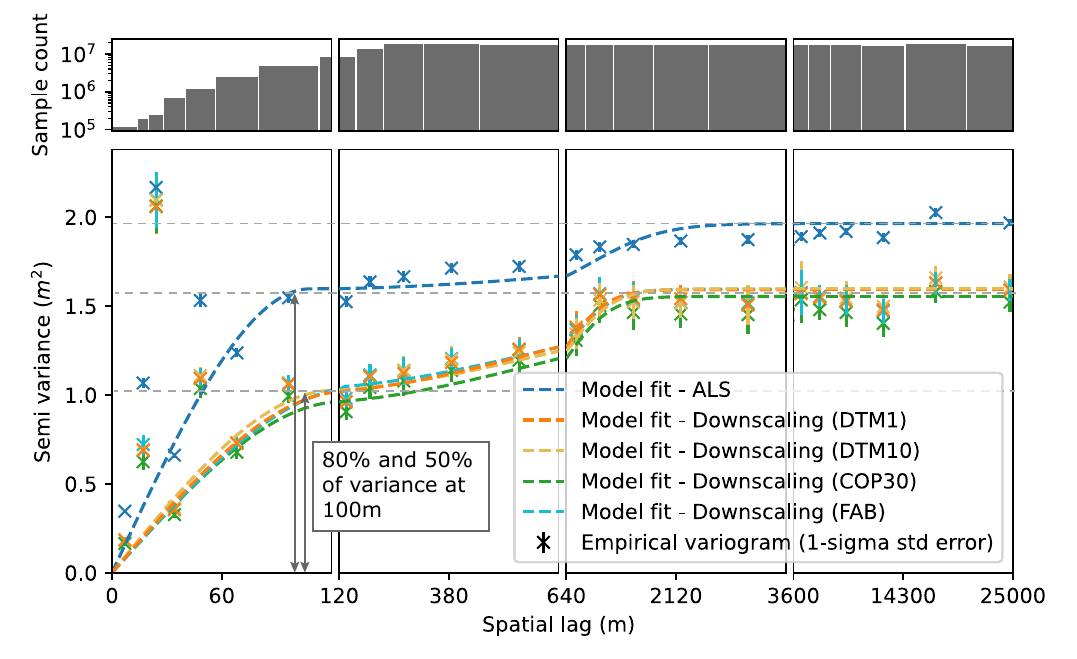}
}
\caption{\label{fig-variogram}Spatial correlation from semivariograms based on six ALS validation strips (2008) and
downscaling model outputs at 10 m resolution. Note the limited capture of
variance in the transverse direction due to the ALS strips'
constrained width (500m). The vertical bar on crosses represents one standard deviation where data are sufficient to compute the statistic.}
\end{figure}

Figure~\ref{fig-snow_presence_2020} provides a visual comparison between the downscaling model output and Sentinel-2 imagery for June 2020 in the lake Møsvatn area (Figure \ref{fig-study_area}). The figure shows the model's accuracy in predicting the occurrence of snow during the rapid melt period. The downscaled snow map from our model visually shows a high level of agreement with the satellite snow extent on June 24, 2020, capturing the remaining snow patches aligning with topographical features (Figure~\ref{fig-snow_presence_2020} c,d). Panel (e) and (f) show that east-facing slopes retain more snow compared to west-facing slopes, a pattern that our model successfully captures. Closer inspection reveals that the model retains a thin snow layer in most areas as well as minor discrepancies in snow distribution. Some of these differences can be explained by the binary nature of Sentinel-2 snow cover and a temporal mismatch, with the Sentinel-2 image taken on a specific day and the model output representing a monthly mean snow depth for June. In particular, the model tends to systematically overestimate snow coverage for specific terrain (indicated by the white circle), such as steep slopes, and underestimates snow coverage for certain terrain features (white circle). To express the match quantitatively, we converted the model output to snow occurrence/probability using a threshold of 0.3 m for full snow cover, and the Sentinel-2 data to a binary reference by computing the Normalised Snow Difference Index (using a threshold of 0.42, as recommended by the Sentinel-2 data provider Copernicus). For the area shown in Figure ~\ref{fig-snow_presence_2020} (extent of panels a,b), we receive a Matthews Correlation Coefficient of 0.35 and an Area Under the Curve (AUC) value of 0.76, both indicating acceptable model performance. For a dynamic view of snow occurrence variability, please refer to the video in the supplementary materials.

\hypertarget{interpreting-the-downscaling-model}{%
\subsection{SHAP analysis and variogram assessment}\label{interpreting-the-downscaling-model}}

The most informative relationships between input features and predicted subgrid variability are listed in descending order.  Elevation (\emph{h\_te\_best\_fit}) emerges as the most significant factor, with higher elevations positively influencing subgrid snow depth (Figure~\ref{fig-shap} a).
Snow depth from ERA5 Land also has a high impact, however, both very shallow snow depth and deep localized snow depth exist (Figure~\ref{fig-shap} a,b) where (\emph{sde\_era}) is high, owing to the low spatial resolution of the data not capturing the variable topography.
Notably, relatively thicker snow (indicated by high SHAP values in Figure~\ref{fig-shap} b,c) is estimated by the downscaling model in concave terrain (negative TPI) with an east-facing slope. In contrast, thinner snow (lower SHAP values) is associated with convex (positive TPI)  and west-facing slopes.
The model estimates deeper snow depths for the two northernmost ALS flight strips of the study area than the southern part (Figures~\ref{fig-shap} d, \ref{fig-meso_chart}). 
The positive cumulative wind aspect factor (\emph{wuf\_positive}) contributes to preferential snow accumulation and shows a stronger influence than the negative wind aspect factor (\emph{wf\_negative}, Figure~\ref{fig-shap} f,g). Most wind aspect factor data points have small values indicative of low wind speeds. These show much variability in their SHAP value, indicating unexplainable variance. The SHAP method cannot distinguish contributions from correlated features such as the apparent importance and positive influence of slope, which might partly be caused by correlated features such as elevation and curvature. The phenomenon of having less snow on very steep slopes (e.g.~\textgreater50°) is not observed (Figure~\ref{fig-shap} a), rather, the model associates high slopes with a positive SHAP value that indicates greater snow depths.

The empirical semivariogram in Figure~\ref{fig-variogram}  shows that the downscaling models based on all DEM versions adequately capture the spatial variability and autocorrelation of snow depths in the Hardangervidda area, but suggests stronger spatial dependence at short ranges for ALS data than downscaled snow depths (ca. 80\% and 50\% of the total variance at 100 m lag). The distance of ca.~ 100 m corresponds to the wavelength of depression features commonly observed in the area (Figure~\ref{fig-snow_micro_2008}). 
Overall, the downscaling model exhibits ca. 20\% less variance than the ALS data. At a distance of ca. 1,300 m, the semi-variance for all datasets approaches the sill.  This distance aligns closely with the typical size of hillslopes in this region, as depicted in Figure~\ref{fig-snow_micro_2008}. The interpolated SeNorge product, based on precipitation gradients, temperature--elevation lapse-rates and disregarding wind processes, has a resolution too coarse to resolve the hillslope-scale snow depth variability.

\hypertarget{discussion}{%
\section{Discussion}\label{discussion}}

\hypertarget{sec-retrieval}{%
\subsection{Bias in elevation datasets}\label{sec-retrieval}}

This study shows that ICESat2 ATL08 data is capable to retrieve snow depths, thus confirming previous findings by other authors \citep{deschamps-berger.etal_2023, enderlin.etal_2022, tian.shan_2021, besso.etal_2024}, while also improving snow depth retreival through co-registeration and rectification of DEMs. Previous studies have in common that retrieved snow depths are biased by remaining discrepancies and errors in the DEM/ICESat-2 data such as a slope-dependent bias. The bias correction introduced in our workflow can reduce errors that depend on first and second-order derivatives (slope/curvature). 
Several studies pointed to ICESat-2 as the primary source of this bias, as ICESat-2 ATL08 data tends to underestimate surface height under certain conditions. \citet{moudry.etal_2022} attributed the error to the presence of clouds and corresponding increasing atmospheric scattering effects leading to an increased photon travel time and, consequently, underestimating terrain height.
Furthermore, the 100 m segment length of ATL08 (e.g. \emph{ h\_te\_mean}  used by \citealt{enderlin.etal_2022}) is considered insufficient for accurately mapping steep and rugged terrain \citep{besso.etal_2024}. Therefore, \citet{enderlin.etal_2022} recommended using ATL08 only in areas with relatively low slopes and sparse vegetation cover. We also observed that bias is higher in the Hardangervidda area than in other, flatter parts of Norway, such as Finnmark (not shown in this paper). In our data sample, a large number of negative snow depth measurements were present before bias correction. Directly excluding these negative snow depths (without further bias correction) would result in unbalanced training samples. A correction solely based on the slope, as proposed by \citet{enderlin.etal_2022} and \citet{tian.shan_2021}, did not remove bias sufficiently for the ICESat-2 measurements to be used as model input for the downscaling. We find resolution-dependent biasing effects for the tested DEMs with regard to curvature that are essential to correct for accurate snow depth retrieval (Supplementary Section B). For Norway, we found that the spatial resolution of the ATL08 data is better than COP30 (30 m) when using \emph{h\_te\_best\_fit} elevation values fitted to the middle 20 m of the 100-m-ATL08 segment (Supplementary Section B).

Also the snow-off DEMs hold potential limitations for ICESat-2-derived snow depths in certain terrain, especially high mountain or Arctic areas where the snow-free season is short and some snow patches remain during summer. Typically, some snow at the highest elevations is of little concern for general-purpose DEM data distributors who need to balance requirements of numerous DEM applications. The DTM1 was acquired during June/July according to metadata, under summer conditions. From comparison with ALS snow-off data (Figure S.6), we find that the Hardangervidda area was not entirely snow-free for any of the four tested DEMs nor the Arctic DEM \citep{porter.etal_2022b}. Remaining snow on the supposedly snow-free reference results in an underestimation of snow depth in parts of these areas, with typical topographic signatures (e.g., shaded depressions). Additionally, in areas with long snow cover duration there are fewer snow-free ICESat-2 segments available. Consequently, the bias correction introduced in this study may not be able to correctly capture and remove this bias, leading to underestimation of high snow depths.  When comparing our downscaled snow depths with the ALS validation data, we find that on the microscale, the
residuals are highly associated with a preferential snow deposition pattern. The systematic underestimation of deep snow depths is even more prominent in the downscaled output if the 
calibration step is not applied (Figures S.8, S.9). The bias correction and calibration steps only indirectly correct for the bias effect of remaining snow patches on retrieved/downscaled snow depths. They also address other bias effects (see section \ref{challenges-of-downscaling-snow-depth}). Future work should thus include a targeted correction of the DEM data for remaining snow patches. This would improve our results and potentially reduce the need for the current calibration step.

\citet{enderlin.etal_2022} deemed COP30 data too imprecise for estimating snow depth. 
Encouragingly, our results demonstrate that the COP30 and FAB DEMs improved after bias correction, and performed comparably with DTM1 and DTM10 at mesoscale in treeless areas (Figure S.7). We attribute this to the bias correction step in our workflow which successfully removes bias for all four tested DEMs (Supplementary material). We are thus confident that the regression-based bias correction is transferable to other regions for the same (global COP30 and FAB DEMs) or different DEM products. As the training area and sample were very large and diverse in this study (entire Norway), further research on the bias-correction method could examine the applicability to different landscapes/topography or the influence of the size of the area and the ICESat-2 snow-off sample. 
We note that the remaining elevation error of 0.48--0.62~m (NMAD after bias correction) for DTM1 and FAB, respectively, may be acceptable for deep snow or a snowpack highly affected by wind redistribution (\textit{e.g.} this study), but can exceed the thickness of shallow snow packs. The influence of uncertain individual samples is mitigated by the approach of this study, where the ICESat-2 measurements are used as a training sample for a downscaling algorithm rather than on their own. The uncertainty may be reduced for regions with less snow depth variability, lowering de-facto the absolute magnitude of discrepancies between observations and model outputs. Further research is needed to better understand the effect of the snow depth sample uncertainty on the downscaled snow depth maps, in particular the poor representation of extremes. It should also be noted that in this study, terrain parameters used for downscaling are based on the DTM10. The spatial resolution of the COP30 and FAB DEMs is coarser (30m) than our model output resolution (10m) and they were used for snow depth retrieval only, not for downscaling. When applying our workflow to other regions the spatial resolution of the output snow depth maps will therefore be limited by the available DEM.

This study only validates the downscaled snow depths above the tree line and further work is needed to assess the performance of the proposed workflow in forested areas where ALS validation data is available. We expect that more, or different predictors may be needed to capture information on snow-canopy processes \citep{mazzotti.etal_2023} influencing the accumulation and ablation of snow. High-resolution lidar data allows for the removal of vegetation, but global-scale DEMs are usually based on photogrammetry or radar interferometry, which may not exclude vegetation cover. There are efforts to create global-scale vegetation-free DEMs like the FAB DEM included in this study, and further developments are expected from machine-learning algorithms. Currently, for the high elevation accuracy required for snow depth retrieval, we recommend to be cautious about using COP30 and FAB in forested areas, as vegetation is hardly (fully) removed.

\hypertarget{challenges-of-downscaling-snow-depth}{%
\subsection{Snow depth downscaling}\label{challenges-of-downscaling-snow-depth}}

Tree-structure models perform well in capturing nonlinear relationships. The model used in this study provides reasonable predictions for snow depth. However, unless calibration is applied, our approach leaves a significant amount of variability unaccounted for, particularly for periods/areas with deep snow depths as observed in the time series analysis (Figure~\ref{fig-snow_weatherstation}) or in the snow depth maps (Figures S.8, S.9). The underestimation of spatial variability arises from several factors: i) the regression model's inherent limitations, ii) a likely imbalance of the training samples, and iii) the limitation of the predictors to capture physical processes driving the redistribution of snow. 

Inherently (i), deterministic regression models primarily yield conditional means and may not adequately represent extreme events, especially when key features do not fully explain variations. This often results in underfitting as the models strive to minimize overall prediction errors. Additionally (ii), the spatial distribution of snow depth is inherently scale-dependent \citep{melvold.skaugen_2013, mott.etal_2018}, meaning a sensor's spatial resolution is critical in determining captured variability. Our training samples, derived from the ICESat-2 ATL08 elevation product, correspond to an area of ca.~20 m × 15 m, which is better than previous study cases but still averages out for finer-scale variability. ICESat-2's sparse spatial sampling pattern with several kilometres gap between sampled snow depth profiles, combined with a 92-day revisit period of the satellite and possible cloud cover leads to substantial gaps in seasonal data coverage. Data points with high snow depth measurements are thus likely to be relatively few. ICESat-2's elevation data, derived from a ground-finding algorithm that use a probability distribution function (PDF) of reflected photons \citep{neuenschwander.etal_2022}, can vary in quality depending on the terrain. In rugged terrain where snow depth is typically the deepest, noise or unbalanced sampling can lead to misrepresentation of extreme values. 

The shortcomings of the model and data sample could be mitigated if model predictors were able to fully capture spatial variability of snow depth and its underlying physical processes (iii), which is not the case. ERA5 Land input snow depth values may provide incorrect prior information as the dataset seems to contain interpolation effects that are possibly a downscaling artefact from the production of ERA5 Land from the coarser ERA5 data \citep{hersbach.etal_2020}. 
For example, snow depth values are much higher for ERA5 Land grid cells on or adjacent to large glaciers like the Jostedalsbreen ice cap in Southern Norway, and lower for grid cells on or adjacent to large water area. Consequently, although we have excluded all glaciers and lakes, neighbouring ERA5 Land grid cells still contain these bias effects. There, the predicted subgrid variability is not solely a reflection of inherent snow conditions but is significantly influenced by this spatially-dependent bias that the model needs to detect and correct for based on spatial information, \textit{i.e.} coordinate features (N, E). This highlights the importance of dense, localized sampling of ICESat-2 snow depths, which may be challenging for smaller study areas and for lower latitudes, given ICESat-2's sparse sampling pattern. By expanding the training area, as done in this study, the model benefits from more samples but might also learn a more generalized, averaged representation of snow depth.

Among the predictor features, elevation emerges as the most significant, as expected from the temperature--elevation relationship, orographic and precipitation shadowing effects \citep{mott.etal_2018, parr.etal_2020} where also aspect plays an important role. The TPI at a 90-meter scale is a strong predictor (Figure \ref{fig-shap}) and corresponds to the size of terrain features in our study area. However, these predictors represent general patterns only. They are not sufficient to represent the locally snow depth variability. Here, the cumulative effect of multiple snow transport events over an entire winter season plays a crucial role. By refining the wind-aspect factor \citep{bennett.etal_2022, dvornikov.etal_2015} into new cumulative values to describe wind redistribution (i.e. accumulation or erosion), our model is able to explain a considerable amount of the observed preferential snow deposition (Figure \ref{fig-shap} g). Erosion is not well captured (Figure \ref{fig-shap} f) by the downscaling model. There is potential to improve wind-related predictors, as the spatial-temporal resolution of wind fields from ERA5 Land monthly data is likely insufficient for capturing complex wind-topography interactions. Downscaling of wind fields \citep{fiddes.etal_2022, letoumelin2023} could enhance the model's ability to capture wind-driven variability.  Additionally, the model does not account for the impact of wind on the energy balance at the snow-atmosphere interface, involving sensible and latent heat exchanges \citep{mott.etal_2017}, a factor crucial to the dynamics of a snowpack. The choice of predictors in this study is based on our expertise and corresponds to commonly used predictors in statistical modelling of snow distribution. This choice inherently introduces bias as the predictors are not able to explain the entire variability. Future research could benefit from incorporating more features that are related to physical snow processes or the integration of snow melt-out information from fractional snow cover data from optical satellite imagery \citep{gascoin.etal_2019} commonly used in snow data assimilation schemes with process-based models \citep{margulis.etal_2016, mazzolini.etal_2024}. Used as an exploratory tool, the model might assist in learning what processes are at play and how to best represent them statistically, ideally leading to a simplification of the workflow.

The calibration step uses quantile mapping based on the observed cumulative snow depth distribution function during 2009 peak snow conditions in the Hardangervidda area. Our selected control points cover the entire study area (\textasciitilde1.5 points per square kilometre), ensuring a comprehensive correction across all regions analyzed. While greatly improving our results, especially for the reproduction of extremes values, the need of a scaling calibration is a disadvantage for an extrapolation of our method to global applications. The slight underestimation of snow depths for the presented western sub-region in the validation area (Figure \ref{fig-snow_micro_2008}) suggests that the single scaling function obtained from the Hardangervidda area does not fit the sub-region perfectly, and a locally adjusted scaling might result in a better fit. 

Future research is needed to determine optimal training/model area sizes or transferable scaling calibration functions for areas with different topography and climate. However, extensive, other publicly available snow depth reference datasets like the Hardangervidda ALS dataset are currently not available for Norway and are generally very rare. In the western United States, the Airborne Snow Observatory data \citep[ASO;][]{painter.etal_2016} includes multi-temporal snow depth maps of several large catchments that could be suitable to examine the local dependency of our downscaling approach and further develop the method. Similarly, the lidar dataset collected in Northern Alaska by \cite{parr.etal_2020} could be used to train for an Arctic snowpack. As an alternative to using extensive validation data for model result calibration, future research could explore the use of small representative snow depth distributions such as strategically chosen, smaller snow depth maps from drone-based surveys, data from snow courses, or snow depth measurements from meteorological stations.

\hypertarget{subgrid-variability-in-the-validation-area}{%
\subsection{Validation} 
\label{subgrid-variability-in-the-validation-area}}

Snow depth distribution in alpine environment is influenced by distinct processes occurring at various spatial and temporal scales. Our model demonstrates high fidelity of snow distribution patterns even at the microscale  (Figure~\ref{fig-snow_micro_2008} c), though with lower performance metrics compared to the mesoscale. Our findings complement the insights from \citet{mott.etal_2018}, which recommend a resolution finer than 50 meters to capture the spatial variability inherent to wind-driven processes. For example, \citet{trujillo.etal_2007, mott.etal_2011, mendoza.etal_2020a} report a distinct 'scale break' at around 100 m, with a stronger spatial autocorrelation of snow depths below the scale break than beyond. Such spatial correlation can be attributed to the wavelength of wind-driven snow accumulation features \citep{mott.etal_2018}.  
The ALS data and our model output identified the presence of scale breaks and dependence on terrain features (95 m accounting for 79\% variance for ALS data and 108 m for 60\% variance for our model, Figure \ref{fig-variogram}), demonstrating the effectiveness of our scheme in capturing a high level of spatial details and the impact of landscape patterns. Notably, we also found a scale break at 1,300 m by fitting the empirical variograms at two different ranges, which corresponds to the wavelength of ridges in this area (Figure \ref{fig-variogram}). These scale breaks underscore the critical resolution and minimal survey scope required to accurately capture spatial variability over the hillslope. Such insights are pivotal for designing snow surveys or evaluating snow models/products.

RMSE or \(R2\) are standard metrics for evaluating regression model performance, focusing primarily on overall accuracy and fit. However, they may not adequately capture the variability and structural patterns in snow depth distribution, especially at the microscale.  Figure~\ref{fig-snow_micro_2008}  and \ref{fig-snow_weatherstation} show high signal fidelity (high Spearman rank correlation) however with poor \(R2\). To enhance the evaluation of regression outcomes for snow depth subgrid variability, it is essential to consider the scales involved and incorporate additional metrics. As an alternative to the Spearman rank correlation used in this study, structural similarity index measure (SSIM, \citealt{parr.etal_2020}) might be a suitable metric to capture relative similarities.

\hypertarget{application}{%
\subsection{Application}\label{application}}

Current limitations of ICESat-2 data, including the month-long data availability lag and significant spatial and temporal gaps, restrict its utility  \citep{deschamps-berger.etal_2023}. Spatio-temporal propagation of ICESat-2 snow depths by use of modelling is thus the logical next step to produce timeseries of snow depth maps and make this data useful for wider applications. Hereby, standard statistical models that solely depend on topographic features for snow spatial distribution fall short on temporal variability. Our workflow resolves many of these constraints by incorporating temporal variability from ERA5 Land. Other emerging approaches to create snow depth maps from ICESat-2 data include combining ICESat-2 snow depths with snow cover information and a process-based model within a data assimilation framework (\citealt{mazzolini.etal_2024}, using higher-resolution snow depths from ATL03 data) or a combination of deep learning and probabilistic data assimilation methods (\citealt{Guidicelli.etal_2023}, using snow depth tracks synthesised from snow depth maps to mimic ICESat-2 snow depths). Both approaches yield promising results and also use ERA5 Land data as forcing, thus in principle globally applicable, but focusing on smaller catchments and coarser spatial resolution than presented in this work. Future research could focus on comparing or merging these different approaches. 

Notably, we find that the performance of aggregated downscaled snow depth is not sensitive to the absolute precision of the reference DEMs (Supplementary Section D), indicating that our approach is not limited to areas with high-resolution DEMs but can be applied using global DEMs in treeless conditions globally. However, careful consideration is required when applying this workflow to new study areas or applications: 
\begin{enumerate}
    \item Global DEMs like the Copernicus GLO-30 or FABDEM are a patchwork of data from satellite-derived DEMs and local/national data sources and may not have equally good quality in different regions,
    \item  Care has to be taken to ensure including enough ICESat-2 samples in/around the target area. Areas further south will have larger coverage gaps due to ICESat-2's sampling pattern that is densest at the poles. Upcoming ICESat-2 data expected for several more years will mitigate this
    \item  Applying calibration with the most representative field survey
    \item  Accounting for regional climatic variability that some area are heavily influenced by specific weather events or patterns.
\end{enumerate}

The workflow described herein is applicable to a range of applications in need to assess snow distribution patterns where sparse snow observations exist. For example, our workflow can (1) mitigate data gaps in remote areas spatially and temporally, e.g. by interpolating and extrapolating weather station observations (Figure  \ref{fig-snow_micro_2008}, \ref{fig-snow_weatherstation}), (2) help to plan and reduce cost of new snow surveys (e.g coordinating surveys with ICESat-2 coverage), (3) improve local estimates of snow water equivalent in large alpine watershed, and (4) be combined with a classical snow modeling approach. Suppose a snow model provides not just average value snow depth, but also a probability density distribution representing the subgrid variability \citep{gisnas.etal_2016}, such a curve can then be translated into a snow depth map using our snow downscaling scheme. 

In the current implementation, we observe some limitations in extrapolating the method in time, or simulating timeseries for a single point. 
First, our current implementation assumes stationarity of ERA5 Land data. For instance, \cite{tc-17-5007-2023} found discontinuities in the ERA5 Land snow products around 2004 due to the introduction of new satellite products in the assimilation scheme. Other stationarity issues may arise from the method used to generate ERA5 Land with successive spin-up periods \citep[1949, 1981, 2001; ][]{essd-13-4349-2021} or due to changed climate conditions in the future. An alternative to expand the temporal extrapolation and improve stationarity issues would be to employ an energy-mass balance snow model like Crocus or Snowpack \citep{gmd-5-773-2012, BARTELT2002123} using ERA5 as forcing. Such snow reanalysis approaches were found to be better at estimating snow water equivalent (SWE) globally than methods based on passive microwave observations \citep{tc-14-1579-2020}. Our statistical approach performs well compared to the seNorge product, but there is currently no physically-based snow reanalysis available at a comparable spatial resolution in Norway. Any subsequent application of the proposed approach will require careful assessment of uncertainties and the quality of input data, including ERA5 Land stationarity. Second, simulating snow depth at the point scale as shown in section \ref{temporal-variability-of-snow-depth} shows poor agreement with station measurements (Table \ref{tab:station-table}), and the application of a simple correction factor to ERA5 Land timeseries would likely yield   better results. This is partly due to various local processes that may affect snow distribution not captured by the model (\textit{e.g.} localised wind field in respect to terrain, radiation, etc.) or locally disturbing the weather station, but also the model capabilities and design with mostly spatially constant input features. Further studies may consider adding predictor features that consider time explicitly or in a summarised way over the course of the snow season, such as a radiation or heat index \citep{cristea.etal_2017} or annual snow melt-out dates from fractional snow cover data. 

Given careful calibration and information on snow density, the presented cost-efficient subgrid parameterization for snow depth could be used to estimate SWE or to correct precipitation/snowfall bias in snow models \citep{girotto.etal_2024}, to eventually serving as input for discharge modeling \citep{helbig.vanherwijnen_2017}. Microscale snow depth maps could benefit studies of ecosystems in snow-covered regions, from habitat availability for wildlife \citep{LISTON2016114} to plant phenology under the snowpack \citep{niittynen2018}. Snow depth parametrizations/distributions at the mesoscale for large areas or detailed snow depth maps at the microscale are crucial for local permafrost studies \citep{gisnas.etal_2016}.

\hypertarget{conclusion}{%
\section{Conclusion}\label{conclusion}}

This study introduces a workflow for snow depth retrieval from ICESat-2 ATL08 and DEM data to downscale ERA5 Land snow depth data using XGBoost tree-structure machine learning models. The two datasets, ICESat-2 and ERA5 Land, have complementary resolutions in space and time that allow for the generation of accurate monthly snow depth maps at the hillslope scale. Hereby, ERA5 Land data primarily provides the temporal variability, and ICESat-2 the spatial variability in snow depths, propagated in space using terrain features and other relevant predictors to train the downscaling model. Advanced bias correction and calibration are part of the workflow to address inherent systematic errors present in the data and correct for residual bias. 

\begin{enumerate}
\def\labelenumi{(\arabic{enumi})}
\tightlist
\item
  There are few snow depth observations available in remote areas, and to this day, no inexpensive ways to map small-scale variability exist. Here, ICESat-2 ATL08 data as presented in our workflow of co-registration and bias-correction stands out as a valuable data source. 
\item
  The downscaling-calibration scheme's performance to predict peak snow for 6×2 ALS flight strips in the Hardangervidda is very good at mesoscale  (100 x 500 m, \(R2\) values ranging from 0.74 to 0.88). At microscale (10 m), the spatial snow depth pattern is captured very well but absolute values are less so. The model is also able to represent the spatial pattern of snow melt-out during late spring as visible from snow cover satellite data.
\item 
We introduce a new cumulative wind-aspect factor in the downscaling model that estimates snow wind re-distribution from ERA5 Land monthly wind fields in a cumulative way. This factor has a high predictive strength for the spatial distribution of snow depth at micro- and mesoscale in the downscaling model.
\item
 The downscaling model is sensitive to systematic bias in the elevation data, like slope- and curvature-dependent bias, which is more critical in global DEMs. Our bias correction demonstrates significant improvements in such DEMs. Therefore, similar results can be obtained when using the global DEM Copernicus GLO-30 (30 m spatial resolution) compared to the Norwegian national DEM (DTM1) at 1 m spatial resolution. 
  \item
   Post-calibration of downscaling model results is currently necessary to compensate for the model's under-representation of extreme values. This under-representation is likely caused by inherent model behaviour (tending towards average values), remaining bias and the nature of the training data. Biasing factors include persistent snow patches on supposedly snow-free reference DEMs, simplified representation of wind processes, under-sampling and underestimation of high snow depths in ICESat-2 data, and temporal non-stationarity in ERA5 Land data. Future research and more adaptable calibration methods for varied scenarios may improve results and remove the need of post-calibration.
\end{enumerate}

The result of this work is a scalable and explainable downscaling model. It provides a heuristic data-driven solution to model snow depth spatio-temporal variability, especially in mountainous regions. While the validation is specific to Hardangervidda in southern Norway the workflow could be applied to other non-vegetated, snowy regions of the world given an existing local calibration dataset.  

\hypertarget{funding}{%
\section*{Funding}\label{funding}}
\addcontentsline{toc}{section}{Funding}

This work was supported by the SNOWDEPTH project, Research Council of Norway (Grant number 325519) and the research project ClimLAND, an EEA/EU collaboration grant between Norway and Romania 2014-2021, project code RO-NO-2019-0415,1290 contract no. 30/2020. It is a contribution to the strategic research initiative LATICE (University of Oslo, project UiO/GEO103920) and also got some support from the French Ministry of Ecological Transition (convention 261168). 

\hypertarget{credit-authorship-contribution-statement}{%
\section*{Credit authorship contribution
statement}\label{credit-authorship-contribution-statement}}
\addcontentsline{toc}{section}{CRediT authorship contribution statement}

\textbf{Zhihao Liu}: Conceptualization, Methodology, Investigation, Data
Curation, Writing -- Original Draft, Software, Visualization.
\textbf{Désirée Treichler}: Conceptualization, Methodology, Writing -- Review \&
Editing, Resources, Supervision, Funding acquisition, Project
Administration. \textbf{Simon Filhol}: Methodology, Writing -- Review \&Editing.

\hypertarget{declaration-of-competing-interest}{%
\section*{Declaration of Competing
Interest}\label{declaration-of-competing-interest}}
\addcontentsline{toc}{section}{Declaration of Competing Interest}

The authors declare that they have no known competing financial
interests or personal relationships that could have appeared to
influence the work reported in this paper.

\hypertarget{data-availability}{%
\section*{Data availability}\label{data-availability}}
\addcontentsline{toc}{section}{Data availability}

The training dataset, which includes retrieved snow depth data from
ICESat-2 ATL08 for Mainland Norway spanning the period from October 2018 to October 2022, is openly accessible through Zenodo with DOI:\url{10.5281/zenodo.10048875}.
The Gradient Descent Coregistration method is open source at \url{https://github.com/GlacioHack/xdem}. The data processing script can be accessed at \url{https://github.com/liuh886/subgrid\_snow}. 
The validation ALS dataset is from the Norwegian Water Resources and Energy Directorate (NVE) at DOI: \url{10.5281/zenodo.2572731}. 
The rest of the datasets are publicly accessible (data sources available in the text).

\hypertarget{acknowledgements}{%
\section*{Acknowledgements}\label{acknowledgements}}
\addcontentsline{toc}{section}{Acknowledgements}

We express our gratitude to NASA for generously providing free access to
the ICESat-2 data and to ECMWF for the ERA5 Reanalysis dataset. We also
extend our appreciation to the Norwegian mapping agency for granting
access to the national DEM, and to Kjetil Melvold/Thomas Skaugen from the Norwegian Water Resources and Energy Directorate (NVE)  for the valuable ALS
dataset.

\hypertarget{references}{%
\section*{References}\label{references}}
\addcontentsline{toc}{section}{References}

\renewcommand{\bibsection}{}
\bibliography{bibliography_revised.bib}

\clearpage

\appendix
\renewcommand{\appendixname}{Supplement} 



\end{document}